\documentclass[showpacs, pra,onecolumn,preprintnumbers ,amsmath, amssymb, superscriptaddress, aps]{revtex4-2}
\usepackage{color}
\usepackage{amsmath,amssymb}
\usepackage{pifont}
\usepackage{amssymb}  
\usepackage{bbold}
\usepackage{float}
\usepackage{subfloat}
\usepackage[squaren]{SIunits}

\usepackage[caption=false]{subfig}
\usepackage{tikz}
\usepackage{makecell}
\usepackage{subfig}
\usepackage{pifont}   
\usepackage{graphicx} 
\graphicspath{{Figures/}}
\usepackage{dcolumn}  
\usepackage{bm}       
\usepackage{multirow} 
\usepackage{placeins}
\usepackage[colorlinks]{hyperref}
\usepackage{mathtools}
\usepackage{appendix}

\begin{document}
	
	\title{Effect  of strain on tunneling time in graphene magnetic barrier}
	\date{\today}
	
	\author{Youssef Fattasse}
	\affiliation{Laboratory of Theoretical Physics, Faculty of Sciences, Choua\"ib Doukkali University, PO Box 20, 24000 El Jadida, Morocco}
	\author{Miloud Mekkaoui}
	\affiliation{Laboratory of Theoretical Physics, Faculty of Sciences, Choua\"ib Doukkali University, PO Box 20, 24000 El Jadida, Morocco}
	\author{Ahmed Jellal}
	\email{a.jellal@ucd.ac.ma}
	\affiliation{Laboratory of Theoretical Physics, Faculty of Sciences, Choua\"ib Doukkali University, PO Box 20, 24000 El Jadida, Morocco}
	\affiliation{Canadian Quantum  Research Center,
		204-3002 32 Ave Vernon,  BC V1T 2L7,  Canada}
	\author{Abdelhadi Bahaoui}
	\affiliation{Laboratory of Theoretical Physics, Faculty of Sciences, Choua\"ib Doukkali University, PO Box 20, 24000 El Jadida, Morocco}

	
	\begin{abstract}
We solve the Dirac equation in three regions of graphene to get the solutions of the energy spectrum in connection to the strain, energy gap, and magnetic field.   The Goos-Hänchen shifts and group delay time will be obtained by applying the stationary phase approximation after the wave functions at the interfaces have been matched.  Our results suggest that the group delay time is influenced by the presence of strain along the armchair and zigzag directions.  We show that the gate voltage and strain  have the ability to change the group delay from subluminality to superluminality. This may have significant uses in high-speed graphene-based nanoelectronics. 
	
	\end{abstract}

	\maketitle

    \section{ Introduction}
    The last few decades have seen a lot of interest in quantum tunneling. Much of the prior research mainly focused on the time required for a particle to tunnel through a barrier \cite{Martin, Hauge}. Considering their superluminality, the group delay time develops into a significant quantity associated with the dynamic component of the tunneling process. Hartman found that the group delay time for a quantum particle tunneling via a rectangular barrier is independent of the barrier width when the barrier is opaque \cite{Hartman}. This phenomenon, which is frequently referred to as the "Hartman effect" \cite{Hartman, Nimtz}, suggests that, for sufficiently large barriers, the particle's effective group velocity \cite{Olkhovsky} can become supra-luminous, which was theorized and experimentally confirmed in the optical analogy \cite{Chiao}. 
    Enders and Nimtz used a microwave wave-guide with a smaller zone that served as a barrier to waves with frequencies below the cutoff frequency in that region to experimentally demonstrate the Hartman effect for the first time \cite{Enders,Enders93}.  When they evaluated the frequency-dependent phase shift of continuous wave microwaves transmitted by the structure, they discovered that the length of the barrier zone had no bearing on the phase shift.   They also discovered that the transit time $L/c$ for a pulse traveling at light speed $c$ over the same barrier distance $L$ in vacuum was shorter than the observed group delay. Inferring that evanescent wave tunneling is superluminal as a result. 
    
    Graphene, a two-dimensional material made of carbon atoms organized in a honeycomb, is one of the most fascinating systems being realized in current science  \cite{Novoselov, Zhang}. This is particularly true since the electron energy spectrum exhibits linear dispersion as a result of the massless Dirac equation, which possesses band structures without energy gaps and an effective velocity $v_F=10^{6}$ ms$^{-1}$ \cite{Geim, Gusynin}. As a result, the monolayer graphene is useless for microelectronic applications as a result of the zero bandgap. To create a bandgap in graphene, several techniques have been proposed, including chemical functionalization \cite{Elias}, patterning it into nanoribbons  \cite{Han}, and inflicting strain on the graphene sheet \cite{Lee}.
     The Dirac points can be altered by the mechanical deformation of graphene, which gives the Dirac fermions asymmetric effective velocities $v_{x}^{\gamma}\ne v_{y}^{\gamma}$ \cite{Choi,Soodchomshom,Yn}.   On the other hand, in the late 1940s, Artman \cite{Artmann11} theorized an explanation for the Goos-Hänchen (GH) shifts , which was discovered by Hermann Fritz Gustav Goos and Hilda Hänchen \cite{Ann47,Ann49}. Many investigations in graphene-based nanostructures, including single \cite{Chen11}, double barrier \cite{Song12}, and superlattices \cite{Chen13}, have demonstrated that GH shifts can be increased by transmission resonances and controlled by adjusting the electrostatic potential and induced gap.

We look into how strain along the armchair and zigzag directions affects the GH shifts, group delay time, and Hartman effect as they pertain to graphene that has been exposed to a magnetic field, a barrier potential, and a mass term. To find the answers to the energy spectrum, we therefore solve the Dirac equation in three different regions.  We will use the stationary phase approximation to generate the analytical formulas for the group delay and GH shifts after matching the wave functions at the interfaces. The Hartman effect's nature is then clarified after we quantitatively evaluate the impact of strain on the GH shifts while accounting for its lateral contribution.  As a result, we demonstrate that the strain modifies the supraluminal and subluminal group delays and the GH shifts in the armchair direction but has a larger impact in the zigzag direction.

  The paper is structured in the following way: In Section 2, we write the relevant Hamiltonian, establish our theoretical issue, and ascertain the eigenspinors and eigenvalues.  In Section 3, transmission, the group delay, and lateral GH shift are calculated using the boundary conditions and current density.  We numerically investigate our findings in Section 4 under various physical parameter settings.  Finally, we summarize and conclude our work.
 
    \section{ Theoretical model}

\begin{figure}[H]
	\centering
	\subfloat[]{
		\centering
		\includegraphics[width=8cm, height=5cm]{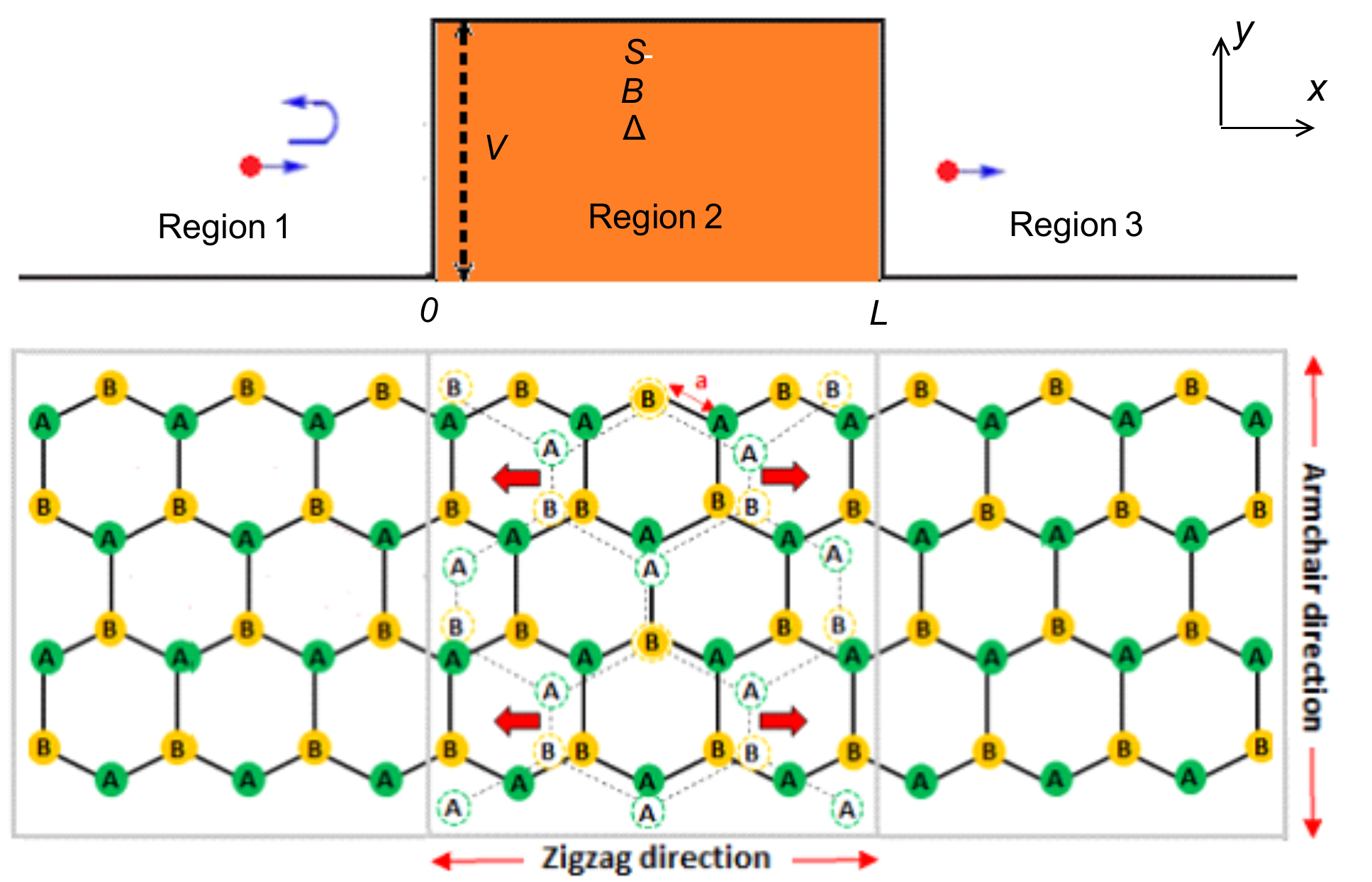}
		\label{db.3}
	}\hspace{1cm}
\subfloat[]{
		\centering
		\includegraphics[width=6cm, height=4cm]{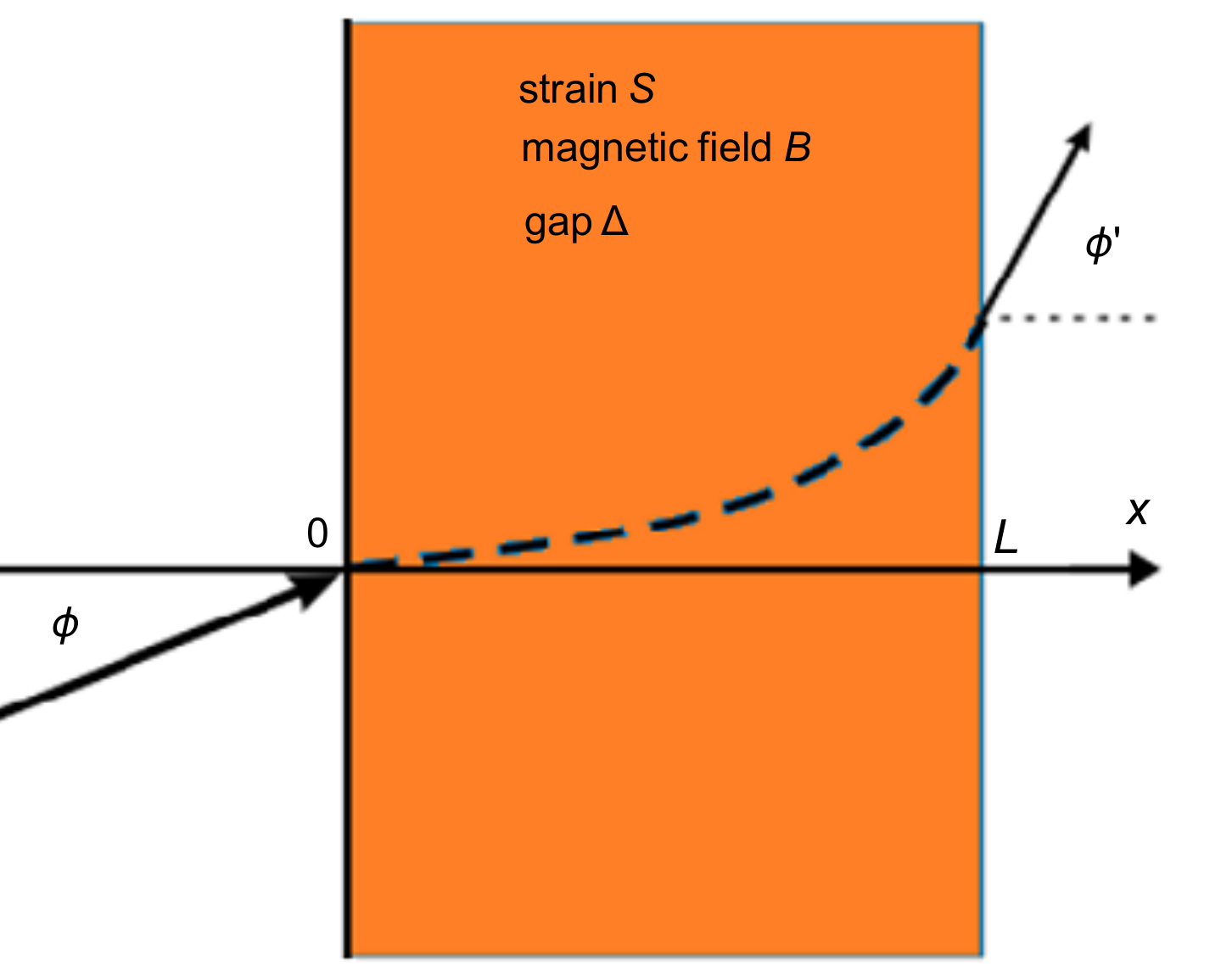}
		\label{db.4}}
	\caption{{(color online) {\color{red}{(a)}}\color{black}{:} Three graphene regions are represented, with the intermediate one subjected to a magnetic field $B$, a gap $\Delta$, and a barrier potential $V$. A strain causing structural deformations of  graphene is applied along the armchair $ (A) $ and zigzag $ (Z) $   directions. 
			{\color{red}{(b)}}\color{black}{:}   Electron trajectory of incident  $\phi$ and  transmitted  $\phi^{\gamma}$ angles, with $\gamma=A, Z$.}}	\label{fig1}
\end{figure}

    As depicted in Fig. \ref{fig1}, a system of massless Dirac fermions is thought to be propagating through a monolayer of graphene. This system is separated into three sections, which are denoted by $j = 1, 2,  3$. There is a square potential, an external magnetic field, and a mass term with tensile strain along the armchair and zigzag directions only in region $2$. Near the Dirac point, we may write the equivalent Hamiltonian for our system as    
    \begin{equation}\label{ham2}
        H_{j}=v_{x}^{\gamma}\sigma_{x}(p_{x}+eA_{x})+v_{y}^{\gamma}\sigma_{y}(p_{y}+eA_{y})+V_{j}(x)\mathbb{I}_2+\Delta\Theta(Lx-x^{2})\sigma_z
    \end{equation}
where the Fermi velocities $v_x^{\gamma}$ and $v_y^{\gamma}$ are tuned by the strain, $\gamma=A$ the strain is in the armchair direction, and $\gamma=Z$ the strain is along zigzag direction, $ (\sigma_{x}, \sigma_{y}, \sigma_z)$  indicate the Pauli matrices, ${\mathbb
	I}_{2}$ represents a $2 \times 2$ unit matrix, $\Theta$ is the Heaviside step function, and $\Delta$ represents a mass term.
We employ a static square potential barrier of the form 
    \begin{equation}
        V_j(x)=
        \left\{%
        \begin{array}{ll}
            V, & \qquad\hbox{$0\leq x\leq L$} \\
            0, & \qquad \hbox{otherwise}. \\
        \end{array}%
        \right.
    \end{equation}
In the barrier zone, an external magnetic field is applied perpendicular to graphene sheet. For further study, we will use the Landau gauge $A(x) = (0,Bx,0)$. 
We consider $v_x^{\gamma}$ and
$v_y^{\gamma}$ as follows 
\begin{align}
	&\label{eqva2}
	v_{x}^{A}= \dfrac{\sqrt{3}}{2\hbar}a(1-\sigma S)+\sqrt{4{t'_{1}}^2-{t'_{3}}^2}  ),\qquad v_{y}^{A}= \dfrac{3}{2\hbar}a(1+S)t'_{3}
\\
&\label{eqva3}
	v_{x}^{Z}= \dfrac{\sqrt{3}}{2\hbar}a(1+S)+\sqrt{4{t'_{1}}^2-{t'_{3}}^2}  ),\qquad  v_{y}^{Z}= \dfrac{3}{2\hbar}a(1-\sigma S)t'_{3}.
\end{align}  
The distance between the closest neighbors in the absence of deformation is given by $ a = 0.142 $ nm, the Poisson ratio is given by $\sigma = 0.165$, and the strain amplitude is given by $S$. Strain affects only the altered hopping integral parameter $t'_i$ in the tight binding approximation, which is caused by stretching or shrinking of the distance vectors between the nearest neighbor carbon atoms \cite{Pereira}. It is represented by an empirical relation. 
\begin{align}\label{eqva6}
	t'_{i}= t_{0} e^{-3.37(|\delta'_{i}|/a-1)},\qquad i=1,2,3
\end{align}
including the jump parameter without deformation, $t_0\approx 2.7$ eV  \cite{Novoselov}. The stress that we applied to the graphene will cause a change in the nearest neighbor jump parameters because such a strain changes the distance of nearest neighbors. Fig. \ref{fig1} shows the graphene atomic structure, with solid and dashed circles signifying sublattices A and B in their undeformed and deformed configurations, respectively. Each circle represents one of the three nearest neighbor vectors, $\delta_i$ and $\delta'_i$. For small strains along the armchair and zigzag directions, $\delta'_{i}$ takes the form
\begin{align}
	&\label{eqva4}
	|\delta_{1}|_{A}= |\delta_{2}|_{A}=a\left(1-\dfrac{3}{4}\sigma S+\dfrac{1}{4}\ S\right),\qquad |\delta'_{3}|_{A}=a(1+S)
\\
&\label{eqva5}
	|\delta'_{1}|_{Z}= |\delta'_{2}|_{Z}=a\left(1+\dfrac{3}{4}\sigma S-\dfrac{1}{4}  S\right),\qquad  |\delta'_{3}|_{Z}=a(1-\sigma S).
\end{align}
By resolving the Dirac equation for the spinor $\Psi_j(x, y)= e^{ik_y y} (\varphi^{+}_{j}(x), \varphi^{-}_{j}(x))^T$, we may find the solutions of the energy spectrum 
    \begin{equation}\label{eqva}
        \left[
    \frac{1}{\hbar} \left(v_{x}^{\gamma}\sigma_{x}({{p}}_{x}+e{{A_{x}}})+v_{y}^{\gamma}\sigma_{y}({{p}}_{y}+e{{A_{y}}})\right)+\tilde{V}_{j}(x)\mathbb{I}_2+\tilde{\Delta}\Theta(Lx-x^{2})\sigma_z\right]\Psi_j(x,y)=
        \epsilon^{\gamma}\Psi_j(x,y)
    \end{equation}
   where  $\epsilon^{\gamma}=\frac{E}{\hbar v_{x}^{\gamma}}$, $\tilde{V}^{\gamma}=\frac{V}{\hbar v_{x}^{\gamma}}$ and
    $\tilde{\Delta}^{\gamma}=\frac{\Delta}{\hbar v_{x}^{\gamma}}$ have been defined, with $ \gamma=A, Z $ refers to armchair and zigzag directions, respectively. \eqref{eqva} will be solved by considering each region of the current system separately.

   In region $1$ $(x<0)$ we solve the eigenvalue equation to end up with the spinor as  the incident and reflected waves, such as
    \begin{eqnarray}\label{555}
        \Psi_1(x,y)=\frac{1}{\sqrt{2}} \begin{pmatrix}
            1\\
            e^{i\phi}
        \end{pmatrix} e^{i (k_x x+k_y y)}+\frac{r}{\sqrt{2}} \begin{pmatrix}
            1\\
            -e^{-i\phi}
        \end{pmatrix} e^{i (-k_x x+k_y y)}
    \end{eqnarray}
where $r$ is the reflection coefficient and  the wave vector component $  k_x $ is
    \begin{equation}\label{kxx}
    k_x=s\sqrt{\epsilon^{2}-k^{2}_{y}}.
    \end{equation}
The incident angle is $\phi=\arctan\left(\frac{k_y}{k_x}\right)$, $\epsilon=\frac{E}{\hbar v_F}$ and $s=\text{sign}(\epsilon)$.
As for  region $3$ $(x > L)$, the  eigenspinor is determined to be 
    \begin{equation}\label{15}
        \Psi_{3}(x,y)= 
        \frac{t}{\sqrt{2}} \begin{pmatrix}
            1\\
            e^{i\phi'}
        \end{pmatrix} e^{i {k}_{x}' x +k_y y}
    \end{equation} 
  including  the transmission coefficient $t$ and the wave vector
    \begin{equation}\label{kpx}
         {k}_{x}'=s\sqrt{\epsilon^2-(k_y+\alpha L)^2}
    \end{equation}
with
 $\alpha=\frac{eB}{\hbar}$. From the two relations
    \begin{equation}
        {k}_{x}'=\epsilon\cos\phi', \qquad (k_y+\alpha L)=\epsilon\sin\phi'
    \end{equation}
we determine the angle 
    \begin{equation}\label{111}
       \phi'=\arcsin\left(\sin\phi+\frac{\alpha L}{\epsilon}\right).
    \end{equation}
    
     The Hamiltonian describing region 2 ($0\leq x\leq L$), can be written as 
    \begin{equation}\label{eq 20}
        H_{2}^{\gamma}=\hbar \left(%
        \begin{array}{cc}
            V^{\gamma +} & -i\sqrt{2v_x^\gamma v_y^\gamma \alpha} a^{\gamma-}\\
            i\sqrt{2v_x^\gamma v_y^\gamma \alpha} a^{\gamma+}  & V^{\gamma -} \\
        \end{array}%
        \right)
    \end{equation}
   where $V^{\gamma \pm}= \tilde{V}^{\gamma}\pm\tilde{\Delta} ^{\gamma}$ has been set. The annihilation  $a^{\gamma-}$ and creation  $ a^{\gamma+} $ operators are introduced as follows:
   \begin{align}
    a^{\gamma\pm}=\frac{1}{\sqrt{2 v_x^{\gamma} v_y^{\gamma} \alpha}}\left[\mp v_x^\gamma\partial_{x}+v_y^\gamma(k_y+x\alpha)\right].
   \end{align}
   The commutation relation is well satisfied $[a^{\gamma- },a^{\gamma+}]=\mathbb{I}$.
    We consider the spinor
$\psi_{2}^{\gamma}(x, y)=e^{ip_{y}y}  (\varphi_{2}^{\gamma +}(x), \varphi_{2}^{\gamma -}(x))^{T}$, $T$ stands for transpose,  and then write
    \begin{equation}\label{eq 23}
        H_{2}^{\gamma}\left(%
        \begin{array}{c}
            \varphi_{2}^{\gamma +} \\
            \varphi_{2}^{\gamma-}\\
        \end{array}%
        \right)=\hbar \epsilon^{\gamma}\left(%
        \begin{array}{c}
            \varphi_{2}^{\gamma +}\\
            \varphi_{2}^{\gamma -}\\
        \end{array}%
        \right)
    \end{equation}
giving rise to the  two coupled equations
    \begin{eqnarray}\label{eq 25}
        &&V^{\gamma +}\varphi_{2}^{\gamma +}-i\sqrt{2v_x^\gamma v_y^\gamma\alpha}a^{\gamma -}\varphi_{2}^{\gamma -}=\epsilon^{\gamma}\varphi_{2}^{\gamma +}\\
        &&\label{eq 26}
        i\sqrt{2v_x^\gamma v_y^\gamma\alpha}a^{\gamma +}\varphi_{2}^{\gamma +} +
        V^{\gamma -}\varphi_{2}^{\gamma -}=\epsilon^{\gamma}\varphi_{2}^{\gamma -}.
    \end{eqnarray}
   For instance, injecting \eqref{eq 26} into \eqref{eq 25} to obtain
    \begin{equation}
        \left(\epsilon^{\gamma}-V^{\gamma +}\right)\left(\epsilon^{\gamma}-V^{\gamma -}\right)\varphi_{2}^{\gamma +}=2v_x^\gamma v_y^\gamma\alpha a^{\gamma -}
        a^{\gamma +}\varphi_{2}^{\gamma +}.
    \end{equation}
This equation represents a harmonic oscillator, and the first component $\varphi_{2}^{\gamma +}$ can be connected to the states $|n-1\rangle$ corresponding to the eigenvalues
    \begin{equation}\label{eq34}
        \tilde{\varepsilon}^{\gamma}=\sqrt{2v_x^\gamma v_y^\gamma\alpha n+(\tilde{\Delta}^{\gamma})^2}, \qquad n\in \mathbb{N}
    \end{equation}
with $\tilde{\varepsilon}^{\gamma}=s'\left(\epsilon^{\gamma}-\tilde{V}^{\gamma}\right)$,
 positive and negative energy solutions are represented by   $s'=\mbox{sign}\left(\epsilon^{\gamma}-\tilde{V}^{\gamma}\right)$. We obtain the second component from  \eqref{eq 26}
    \begin{equation}
        \varphi^{\gamma-}_{2}=s'i\sqrt{\frac{\tilde{\varepsilon}^{\gamma}-s'
                \tilde{\Delta}^{\gamma} }{\tilde{\varepsilon}^{\gamma}+s' \tilde{\Delta}^{\gamma}}} \mid
        n\rangle.
    \end{equation}
Using the parabolic cylinder functions $D_{n}(x)=2^{-\frac{n}{2}}e^{-\frac{x^{2}}{4}}H_{n}\left(\frac{x}{\sqrt{2}}\right)$ to write the solution in region 2
    \begin{eqnarray}
        \Psi_{\sf 2}(x,y) = e^{ik_{y}y}\sum_{\pm}c^{\pm}
        \begin{pmatrix}
            \sqrt{\frac{\tilde{\varepsilon}^{\gamma}+s'\tilde{\Delta}^{\gamma}
                }{\tilde{\varepsilon}^{\gamma}}}
            D_{\left((\tilde{\varepsilon}^{\gamma})^{2}-
            	(\tilde{\Delta}^{\gamma})^{2} \right)/2v_x^\gamma v_y^\gamma\alpha-1}
            \left(\pm \sqrt{\frac{2v_y^\gamma}{v_x^\gamma\alpha}}\left(\alpha x+k_{y}\right)\right) \\
            \frac{ \pm i\sqrt{2v_x^\gamma v_y^\gamma\alpha}}{\sqrt{\tilde{\varepsilon}
                    \left(\tilde{\varepsilon}^{\gamma}+s' \tilde{\Delta}^{\gamma}\right)}}
            D_{\left((\tilde{\varepsilon}^{\gamma})^{2}-(\tilde{\Delta}^{\gamma})^{2}\right)/2v_x^\gamma v_y^\gamma\alpha}
            \left(\pm \sqrt{\frac{2v_y^\gamma}{v_x^\gamma\alpha}}\left(\alpha x+k_{y}\right)\right) \\
        \end{pmatrix}%
       \end{eqnarray}
  where $ H_n(x) $ are Hermite polynomials and $ c^{\pm} $ are two constants.
   Keep in mind that the boundary conditions at interfaces will be used to calculate the transmission $t$ and reflection $r$ coefficients.

    \section{Group delay time}

We look at how strain along zigzag and armchair directions affects the GH shifts and group delay times around a transverse wave vector $k_y $ and incident angle $\phi \in [0, \frac{\pi}{2}]$ meeting  \eqref{111}.  After that, we introduce the critical incident angle $\phi_{c}$, which is determined by   
    \begin{equation}
        \phi_{c}=\arcsin\left(1-\frac{ \alpha L}{\epsilon}\right)
    \end{equation}
and corresponds to $ \phi'=\frac{\pi}{2} $ in \eqref{111}.
We therefore get oscillating guided modes in the case of $\phi< \phi_{c}$ and evanescent wave modes in the case of $\phi> \phi_{c}$. After matching   the eigenspinors at the boundaries $x=0$ and $x=L$ 
    \begin{equation}
            \Psi_1(0,y)=\Psi_{2}(0,y), \qquad
            \Psi_{2}(L,y)=\Psi_{3}(L,y)
    \end{equation} 
we demonstrate the following forms of the transmission and reflection coefficients
    \begin{eqnarray}
        && t^{\gamma}=\frac{2\zeta^{\gamma}_{1}\zeta^{\gamma}_{2}}{\kappa^{\gamma}} (v^{\gamma}_{1L}w^{\gamma}_{2L}+v^{\gamma}_{2L}w^{\gamma}_{1L})\cos{\phi}\\
        &&r^{\gamma}= \frac{\Pi^{\gamma}}{\kappa^{\gamma}}
    \end{eqnarray}
including the quantities
    \begin{eqnarray}
        &&\kappa^{\gamma}=\left(\zeta^{\gamma}_{2}v^{\gamma}_{20}-\zeta^{\gamma}_{1} v^{\gamma}_{10}e^{-i\phi}\right) \left(\zeta^{\gamma}_{1}
        w^{\gamma}_{1L}e^{i\phi'}-\zeta^{\gamma}_{2}w^{\gamma}_{2L}\right)+\left(\zeta^{\gamma}_{1}
        v^{\gamma}_{1L}e^{i\phi'}+\zeta^{\gamma}_{2}v^{\gamma}_{2L}\right)\left(\zeta^{\gamma}_{2}w^{\gamma}_{20}+\zeta^{\gamma}_{1}
        w^{\gamma}_{10}e^{-i\phi}\right)\\
        &&\Pi^{\gamma}=\left(-\zeta^{\gamma}_{2}v^{\gamma}_{20}-\zeta^{\gamma}_{1} v^{\gamma}_{10}e^{i\phi}\right) \left(\zeta^{\gamma}_{1}
        w^{\gamma}_{1L}e^{i\phi'}-\zeta^{\gamma}_{2}w^{\gamma}_{2L}\right)+\left(\zeta^{\gamma}_{1}
        v^{\gamma}_{1L}e^{i\phi'}+\zeta^{\gamma}_{2}v^{\gamma}_{2L}\right)\left(-\zeta^{\gamma}_{2}w^{\gamma}_{20}+\zeta^{\gamma}_{1}
        w^{\gamma}_{10}e^{i\phi}\right) \qquad 
    \end{eqnarray}
    and the following
    shorthand notation
    \begin{eqnarray}
        v^{\gamma}_{1x}&=& D_{\left((\tilde{\varepsilon}^{\gamma})^{2}-(\tilde{\Delta}^{\gamma})^{2}
            \right)/2v_x^\gamma v_y^\gamma\alpha-1}
        \left(\sqrt{\frac{2v_y^\gamma}{v_x^\gamma\alpha}}\left(\alpha x+k_{y}\right)\right),\qquad
        v^{\gamma}_{2x}=D_{\left((\tilde{\varepsilon}^{\gamma})^{2}-(\tilde{\Delta}^{\gamma})^{2}\right)/2v_x^\gamma v_y^\gamma\alpha}
        \left(-\sqrt{\frac{2v_y^\gamma}{v_x^\gamma\alpha}}\left(\alpha x+k_{y}\right)\right)\\
        w^{\gamma}_{1x}&=& D_{\left((\tilde{\varepsilon}^{\gamma})^{2}-(\tilde{\Delta}^{\gamma})^{2} \right)/2v_x^\gamma v_y^\gamma\alpha-1}
        \left(\sqrt{\frac{2v_y^\gamma}{v_x^\gamma\alpha}}\left(\alpha x+k_{y}\right)\right),\qquad
        w^{\gamma}_{2x}=D_{\left((\tilde{\varepsilon}^{\gamma})^{2}-(\tilde{\Delta}^{\gamma})^{2}\right)/2v_x^\gamma v_y^\gamma\alpha}
        \left(-\sqrt{\frac{2v_y^\gamma}{v_x^\gamma\alpha}}(\alpha x+k_{y})\right) \qquad  \\ 
        \zeta^{\gamma}_{1}&=&\sqrt{\frac{\tilde{\varepsilon}^{\gamma}+s'\tilde{\Delta}^{\gamma}
            }{\tilde{\varepsilon}^{\gamma}}}, \qquad
        \zeta^{\gamma}_{2}=\frac{i\sqrt{2v_x^\gamma v_y^\gamma\alpha}}{\sqrt{\tilde{\varepsilon}^{\gamma}
       \left(\tilde{\varepsilon}^{\gamma}+s' \tilde{\Delta}^{\gamma}\right)}}. 
    \end{eqnarray}
   It can be easily shown that
    \begin{equation}\label{pshi}
        t^{\gamma}=|t^{\gamma}|e^{i\varphi^{\gamma}_{t}}, \qquad r^{\gamma}=|r^{\gamma}|e^{i\varphi^{\gamma}_{r}}
    \end{equation}
    in which  $\varphi_{t}^{\gamma}$ and  $\varphi_{r}^{\gamma}$ are the phase shifts. The current density, which is used to determine the transmission and reflection probabilities, is 
    \begin{equation}
        J=ev_F
        \psi^+\sigma_x\psi.
    \end{equation}
This will allow us to find its incident  $J_{\sf in}$, transmitted $J_{\sf tr}$ and reflected $J_{\sf re}$ components. As a result, we obtain the transmission
 $ T^{\gamma}=\frac{|J_{\sf tr}|}{|J_{\sf in}|} $
and reflection $ R^{\gamma}=\frac{|J_{\sf re}|}{|J_{\sf in}|} $ probabilities
     \begin{equation}
     T^{\gamma}=\frac{ {k}_{x}'}{k_x}|t^{\gamma}|^2, \qquad R^{\gamma}=|r^{\gamma}|^2.
     \end{equation}

In what follows, we will look at the resources that can be used to investigate group propagation time in transmission and reflection. 
Notably, a time-space wave packet can represent a finite-pulsed electron beam as a weighted superposition of plane wave spinors. According to \cite{chen08}, the incident, reflected, and transmitted beam waves at the interfaces $ (x = 0, x = L) $ can thus be expressed as double Fourier integrals over $\omega$ and $k_y$. The following integrals are among them:
    \begin{eqnarray}
        &&\label{int1}
        \Phi_{\sf in}(x,y, t)=\iint f(k_y,\omega)\ \Psi_{\sf in} (x,y) \  e^{-i\omega
            t}\ dk_yd\omega\\
   &&\label{int2}
        \Phi_{\sf re}(x,y,t)=\iint  r^{\gamma}f(k_y,\omega)\ \Psi_{\sf re} (x,y) \ e^{-i \omega
            t}\ dk_yd\omega
    \\
    &&\label{int3}
        \Phi_{\sf tr}(x,y,t)=\iint  t^{\gamma}f(k_y,\omega) \ \Psi_{\sf tr} (x,y) \ e^{-i \omega
            t}\ dk_yd\omega
    \end{eqnarray}
in which the spinors $ \Psi_{\sf in}, \Psi_{\sf re} $ and  $ \Psi_{\sf tr} $ are  provided in \eqref{555}
and \eqref{15}.   According to \cite{Beenakker}, the angular spectral distribution is considered to have a Gaussian form $f(k_y,\omega)=w_ye^{-w_{y}^2(k_y-\omega)^2}$ with the half beam width at the waist being $\omega_y$, and the wave frequency is $\omega=E/\hbar$. Injecting \eqref{pshi} into (\ref{int2}-\ref{int3}) to determine  the total phases of the reflected and transmitted wave functions at ($x = 0 $, $x = L $). This process yields to 
    \begin{equation}
    \mathbf{\Phi}^{\gamma}_{r}=\varphi^{\gamma}_{r}+k_yy-\omega t, \qquad \mathbf{\Phi}^{\gamma}_{t}=\varphi^{\gamma}_{t}+(k_y+\alpha L)y-\omega t
    \end{equation}
with $\eta=t, r$ stands for the transmission and reflection.
 Using the stationary phase approximation \cite{Steinberg, Li1}, i.e.  $  \frac{\partial\mathbf{\Phi}^{\gamma}_\eta}{\partial\phi}=0 $ and $  \frac{  \partial\mathbf{\Phi}^{\gamma}_\eta}{\partial\omega}=0 $, we calculate the GH shifts $ S^{\gamma}_{\eta} $  and  group delay time $ \tau^{\gamma}_{\eta} $
    \begin{align}
    &\label{ghss}	S^{\gamma}_{\eta}=- \frac{\partial \varphi^{\gamma}_{\eta}}{\partial
    	k_{y}}\\	
   &     \tau^{\gamma}_{\eta}= \tau^{\varphi^{\gamma}_{\eta}} +
        \tau^{s^{\gamma}_{\eta}}=\frac{\partial \varphi_{\eta}}{\partial
        	\omega}+\left(\frac{\partial k_y}{\partial
        	\omega}\right)S^{\gamma}_{\eta}
    \end{align}
   such that
    $ \tau^{\varphi^{\gamma}_{\eta}}$ indicates  the time derivative of
    phase shifts, whereas  $\tau^{s^{\gamma}_{\eta}}$  is the outcome of the $S^{\gamma}_{\eta}$
    contribution. 
    $\tau_{\eta}{\gamma}$  can be thought of as the average of the group delay times of the two components because the wave function involves a two-component spinor. As a result, we have in phase shifts
    \begin{eqnarray}
     \tau^{\varphi^{\gamma}_{t}}=\hbar \frac{\partial \varphi^{\gamma}_t}{\partial
            E}+\frac{\hbar}{2}\frac{\partial \phi'}{\partial E}, \qquad
        \tau^{\varphi_{r}^{\gamma}}=\hbar \frac{\partial \varphi^{\gamma}_r}{\partial E}
    \end{eqnarray}
    and in GH shifts
    \begin{equation}
        \tau^{s_{t}^{\gamma}}=\frac{\sin\phi}{\upsilon_F}S^{\gamma}_t, \qquad
        \tau^{s_{r}^{\gamma}}=\frac{\sin\phi}{\upsilon_F}S^{\gamma}_r.
   \end{equation}
 The numerical analysis of these findings will enable us to comprehend how strain affects the group delay time of a gap opening in a square potential-scattered graphene magnetic barrier.

   \section{Numerical results}
   
   In the case of normal and oblique incidence, the group delay time $\tau_t^{\gamma}$ and GH shifts $S_t^{\gamma}$ will be examined for electrons passing over a magnetic barrier in strained graphene. The behavior of $\tau_t^{\gamma}/\tau_0$ as a function of different physical parameters is examined to determine whether the Hartman effect is present in the system.  $\tau_0$ is the amount of time it would take a particle to traverse the same distance $L$ without a barrier. 
 Based on (\ref{kpx}) and in the case of normal incidence, i.e., $k_y = 0$,  we show that the magnetic field must meet the following requirement 
   \begin{equation}
   \frac{1}{v_y^\gamma}	\frac{\sqrt{|(E-V)^2-\Delta^2|}}{e L }<B< \frac{E}{eL \upsilon_F}.
   \end{equation}
It is worth noting that increasing $B$ causes $k_x'$ to become imaginary,  leading to the evanescent wave function inside the barrier, as shown by \eqref{kpx}. 
 With these parameters $\phi=0^{\circ}$, $V=100 $ meV, $E = 150 $ meV, $L = 80 $ nm and $S=0.1$, we get the intervals 0.77  T $ < B < $  1.87  T for armchair  and 0.58  T $ < B < $  1.87  T for zigzag directions, where the wave function propagates in the transmission region.    
    
\begin{figure}[H]
   	   	\centering
   	\subfloat[$ S=0, S=0.05, S=0.15 $]{
   		\centering
   		\includegraphics[width=8cm, height=5cm]{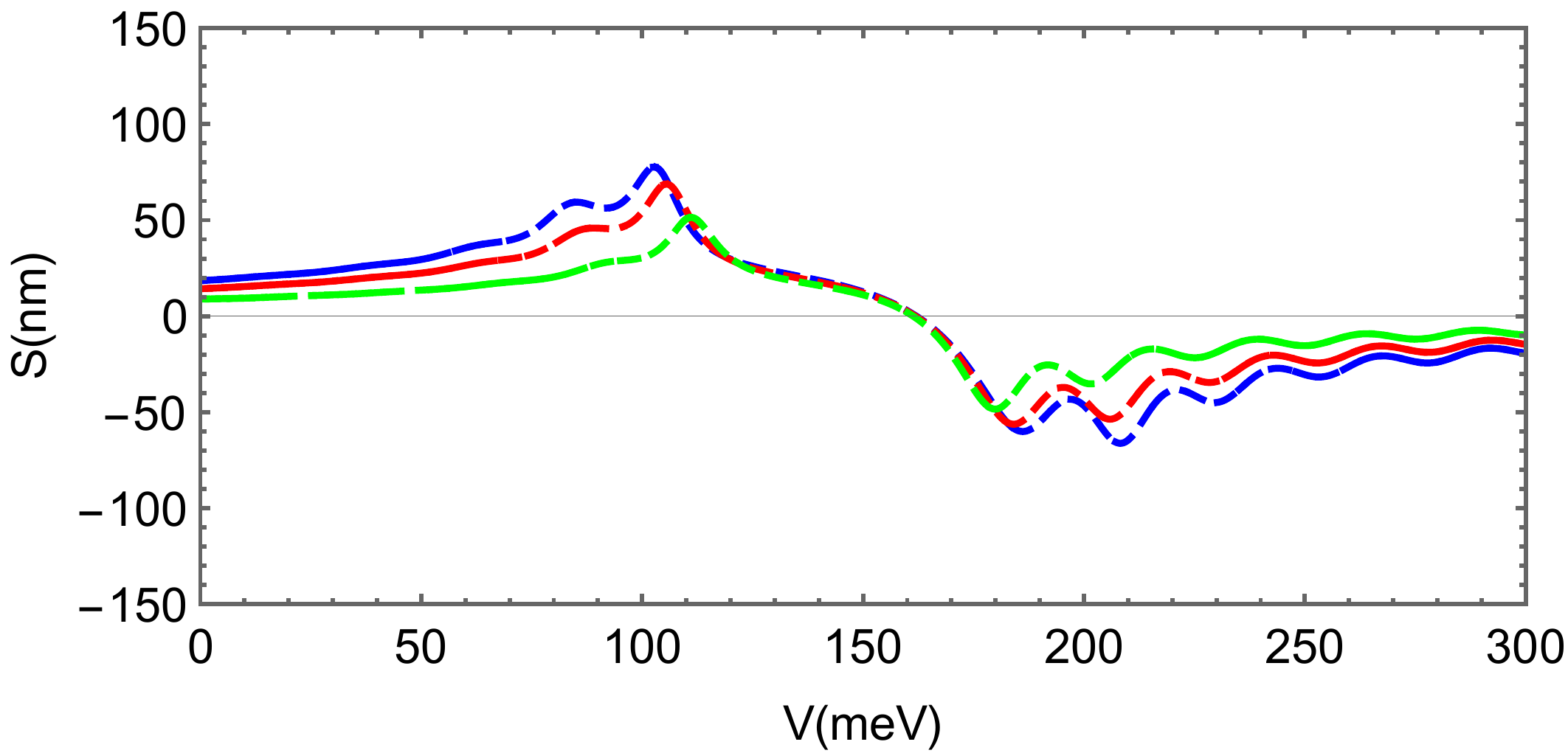}
   		\label{fig2a}
   	}
   \hspace{1cm}
   \subfloat[$ S=0, S=0.05, S=0.15 $]{
   		\centering
   		\includegraphics[width=8cm, height=5cm]{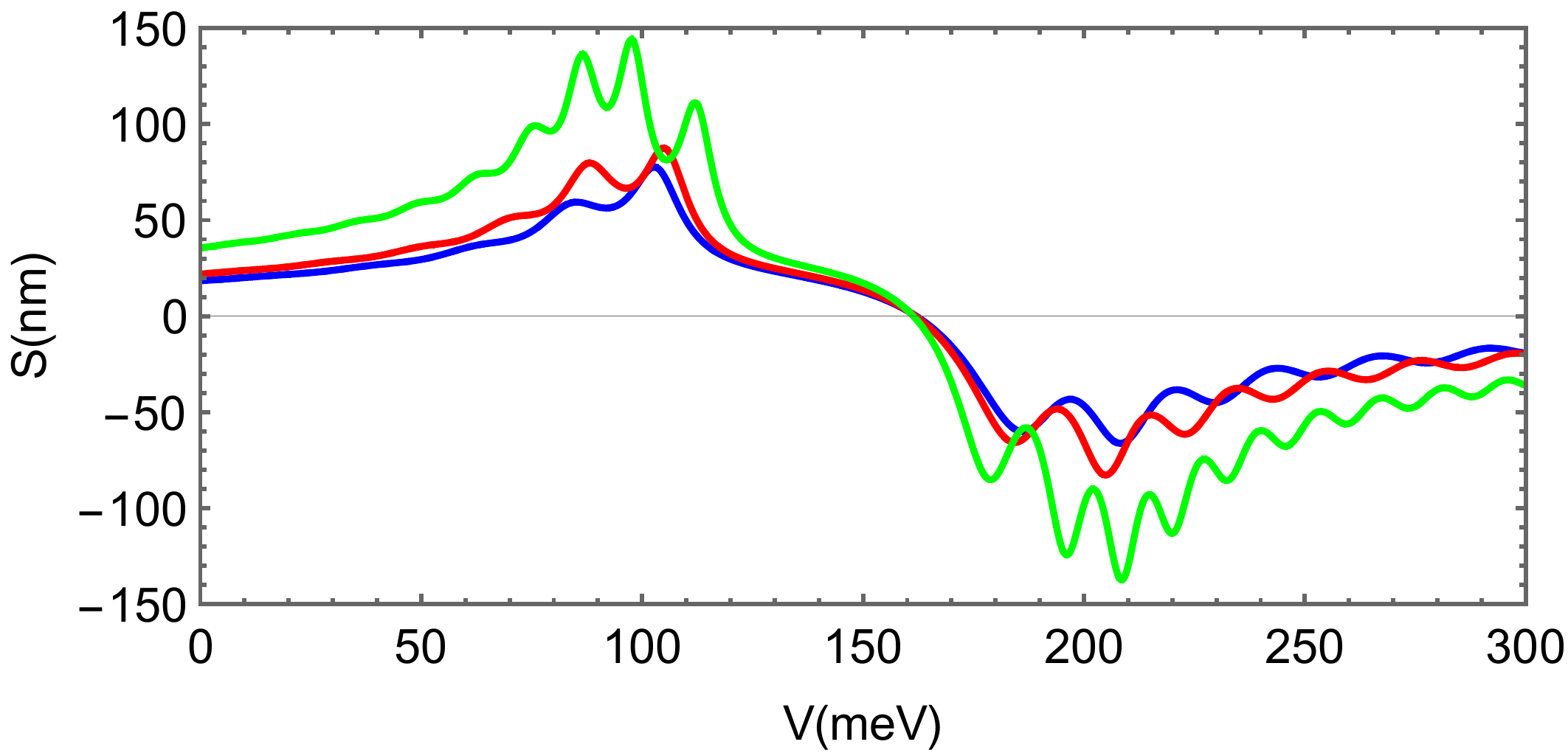}
   		\label{fig2b}}
   	\caption{{(color online) GH shifts and  transmission  versus  the barrier height $V$ for 
   			$S=0$ (blue), $S=0.05$, (red), $S=0.15$, (green)
   			with $E=150$ meV, $\Delta=10$  meV, $B=0.8$ T, $L=80$ nm, $\phi=10^{\circ}$.
   			 \textbf{\color{red}{(a)}}: 	armchair strain direction and
   			\textbf{\color{red}{(b)}}:
   			zigzag strain direction. }}
   	\label{fig2}
   \end{figure}

  The GH shifts in transmission versus the barrier height $V$ are shown in Fig. \ref{fig2} for three strain values: $S=0$ (blue), $S=0.05$ (red), and $S=0.15$ (green), with $E=150$ meV, $\Delta=10 $ meV, $B=0.8$ T, $L=80$ nm, and $\phi=10^{\circ}$. Fig. \ref{fig2a} and  Fig. \ref{fig2b} show the armchair and zigzag cases, respectively.
  The GH shifts change sign near the Dirac point, $V=E$, as can be seen. According to \cite{Chen11}, the change in the sign of the GH shifts results from the transition between the Klein effect ($V>E$) and the classical movement ($V<E$).  
We observe that GH shifts increase as long as $ V $ increases. However, for the case $V>E$, the GH shifts become negative, and when large enough $V$, the GH shifts stabilize and approach zero. Contrary to $ V > E $, as $S$ increases, the GH shifts decrease in the armchair direction but increase in the zigzag direction for $V<E$.  

\begin{figure}[H]
	\centering
	\subfloat[$S=0$, $S=0.05$, $S=0.15$]{
		\centering
		\includegraphics[width=8cm, height=4.8cm]{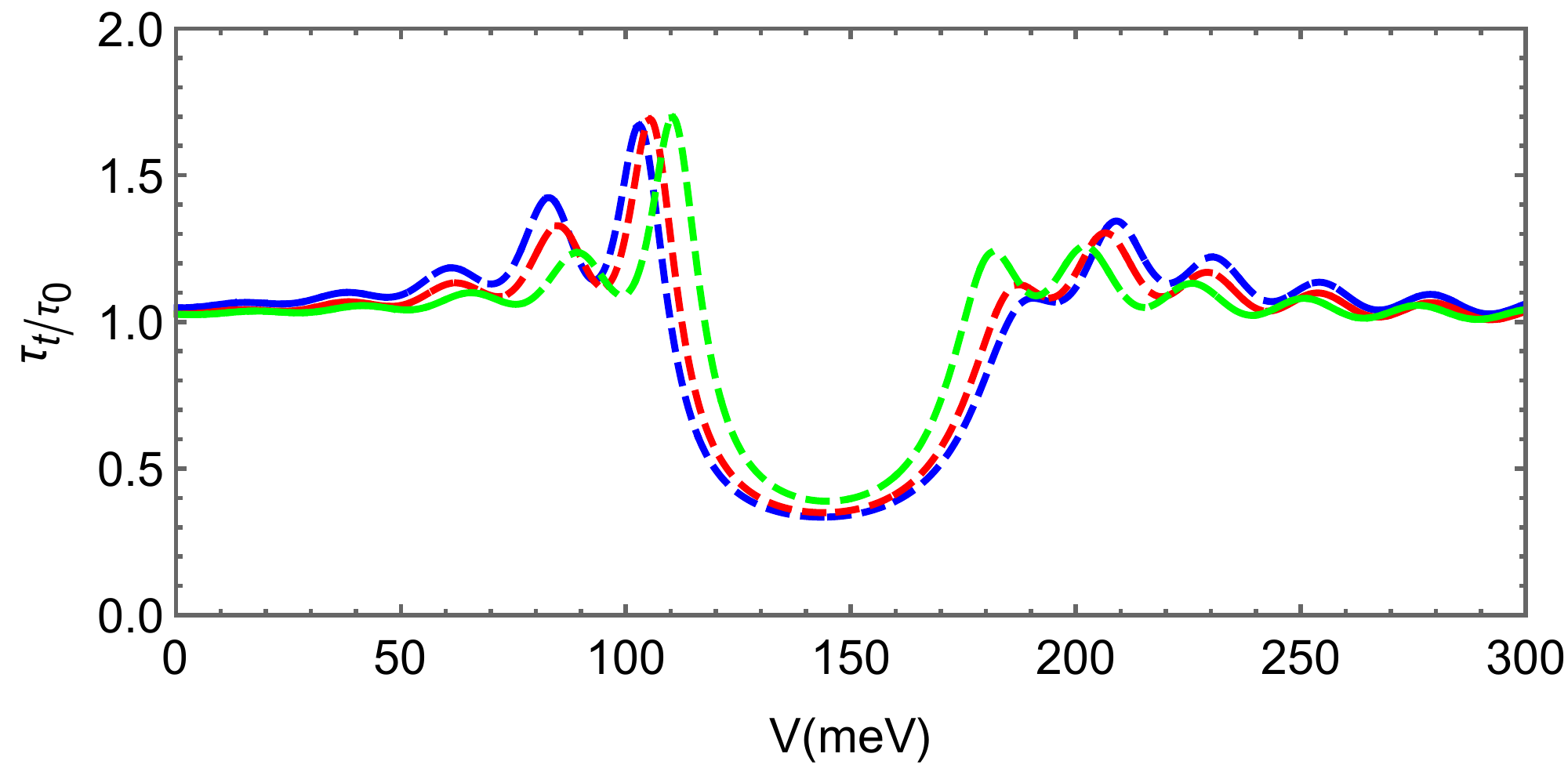}
		\label{fig3a}
	}\hspace{1cm}
\subfloat[$S=0$, $S=0.05$, $S=0.15$]{
		\centering
		\includegraphics[width=8cm, height=4.8cm]{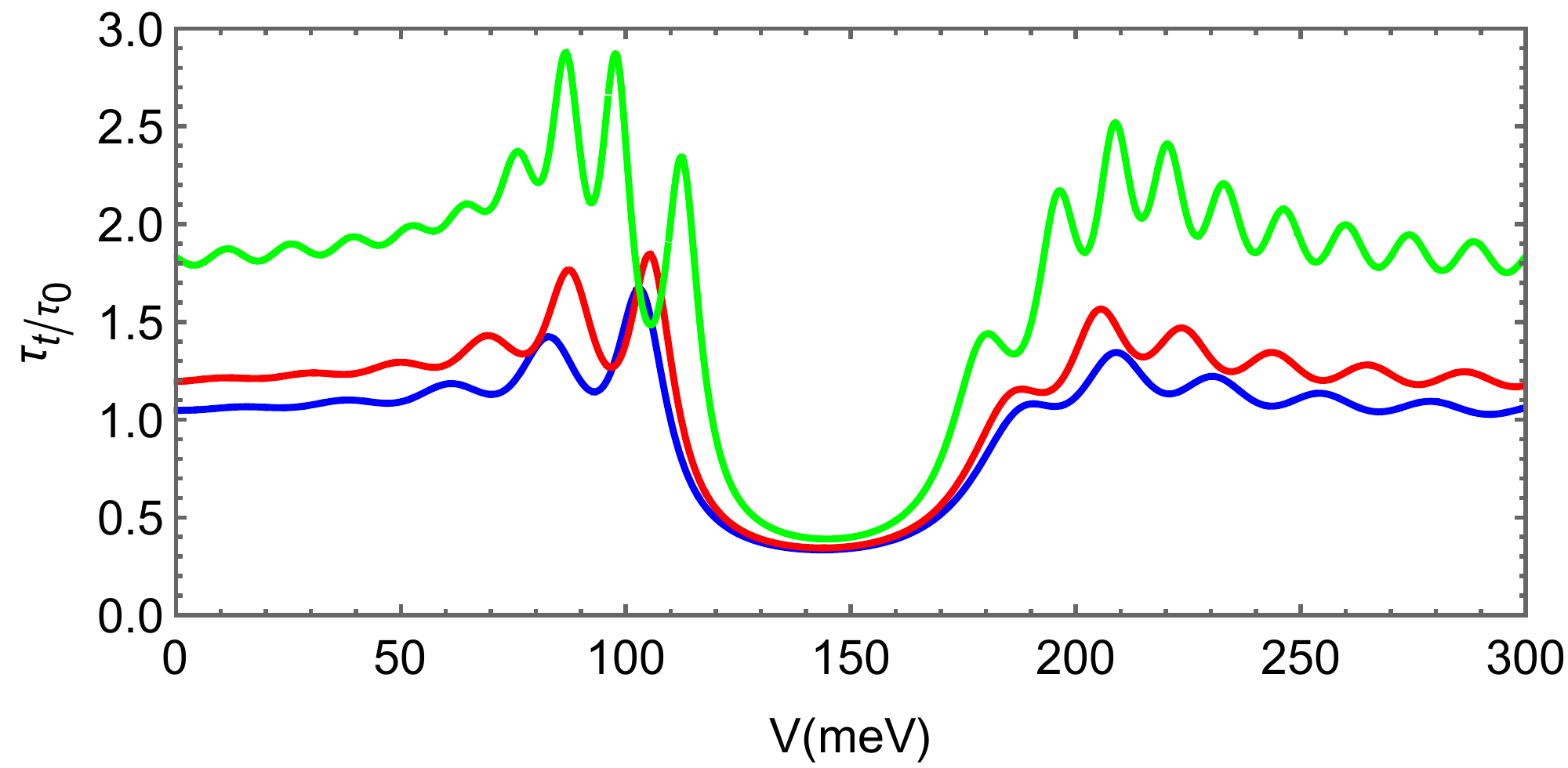}
		\label{fig3b}}\\
		\centering
	\subfloat[$E=150$ meV, $E=170$ meV, $E=180$ meV]{
		\centering
		\includegraphics[width=8cm, height=4.9cm]{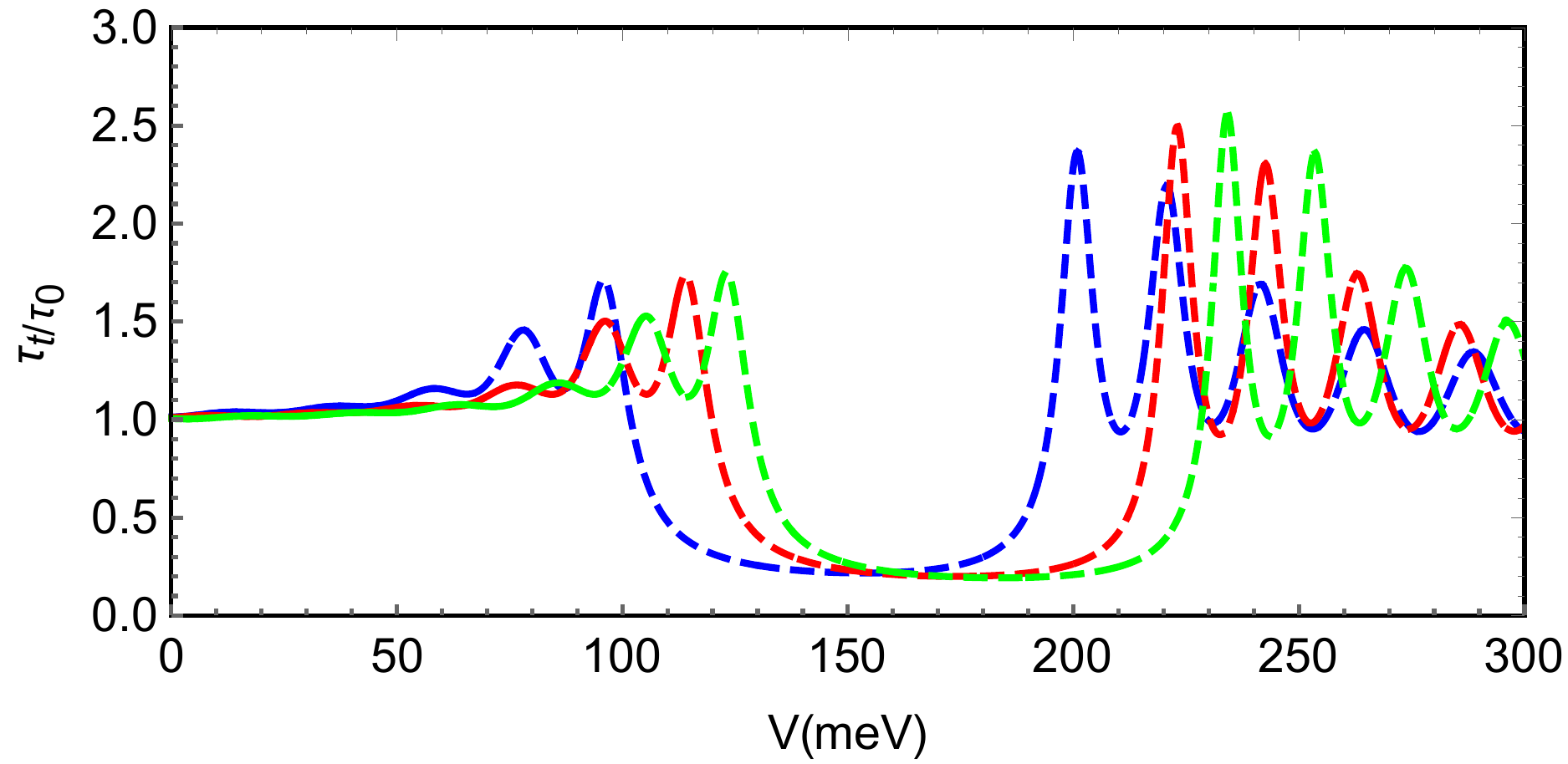}
		\label{fig3c}
	}\hspace{1cm}
\subfloat[$E=150$ meV, $E=170$ meV, $E=180$ meV]{
		\centering
		\includegraphics[width=8cm, height=4.8cm]{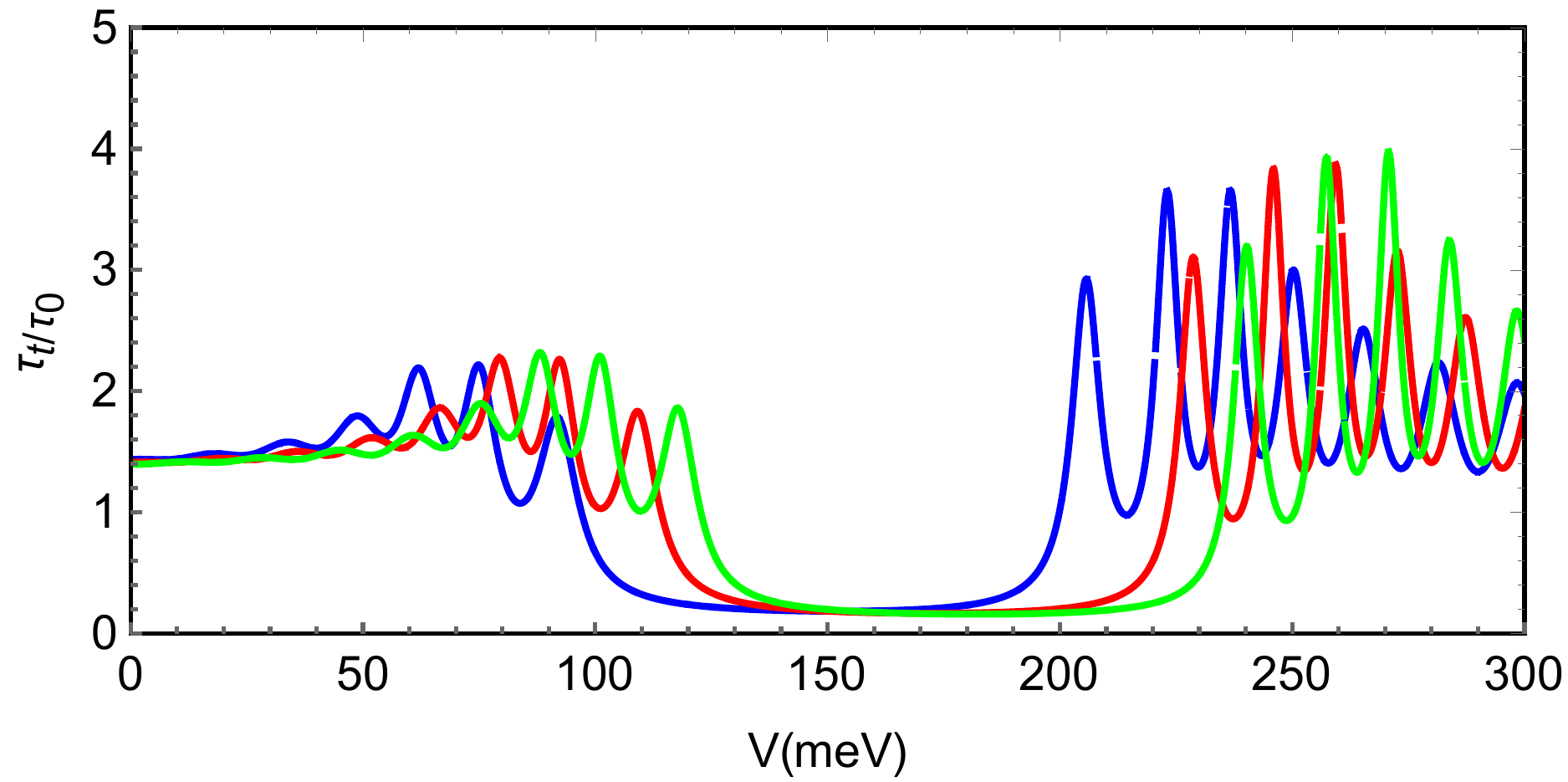}
		\label{fig3d}}\\
	\centering
	\subfloat[$\phi=5^{\circ}$, $\phi=10^{\circ}$, $\phi=15^{\circ}$]{
		\centering
		\includegraphics[width=8cm, height=4.8cm]{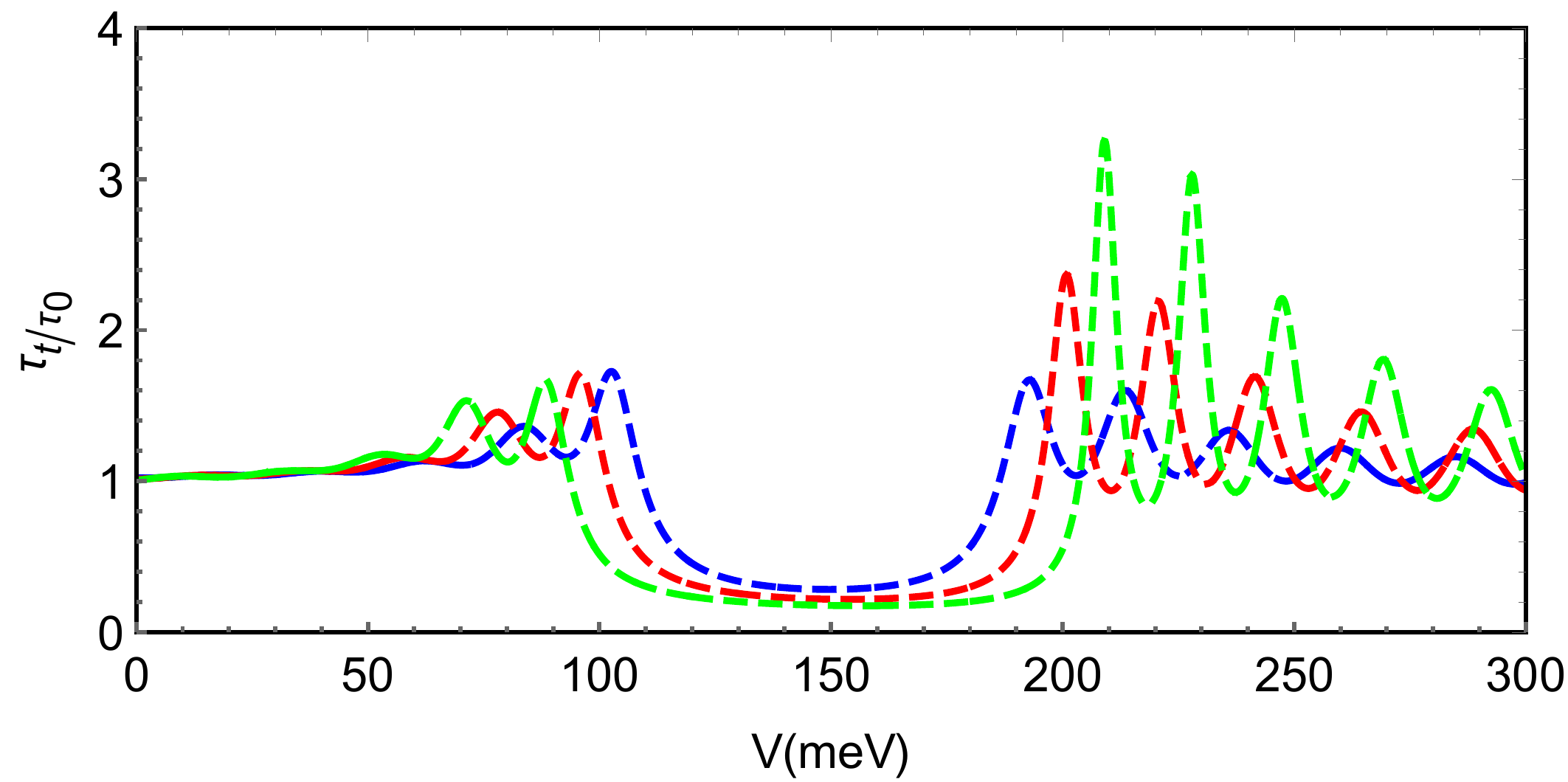}
		\label{fig3e}
	}\hspace{1cm}
\subfloat[$\phi=5^{\circ}$, $\phi=10^{\circ}$, $\phi=15^{\circ}$]{
		\centering
		\includegraphics[width=8cm, height=4.8cm]{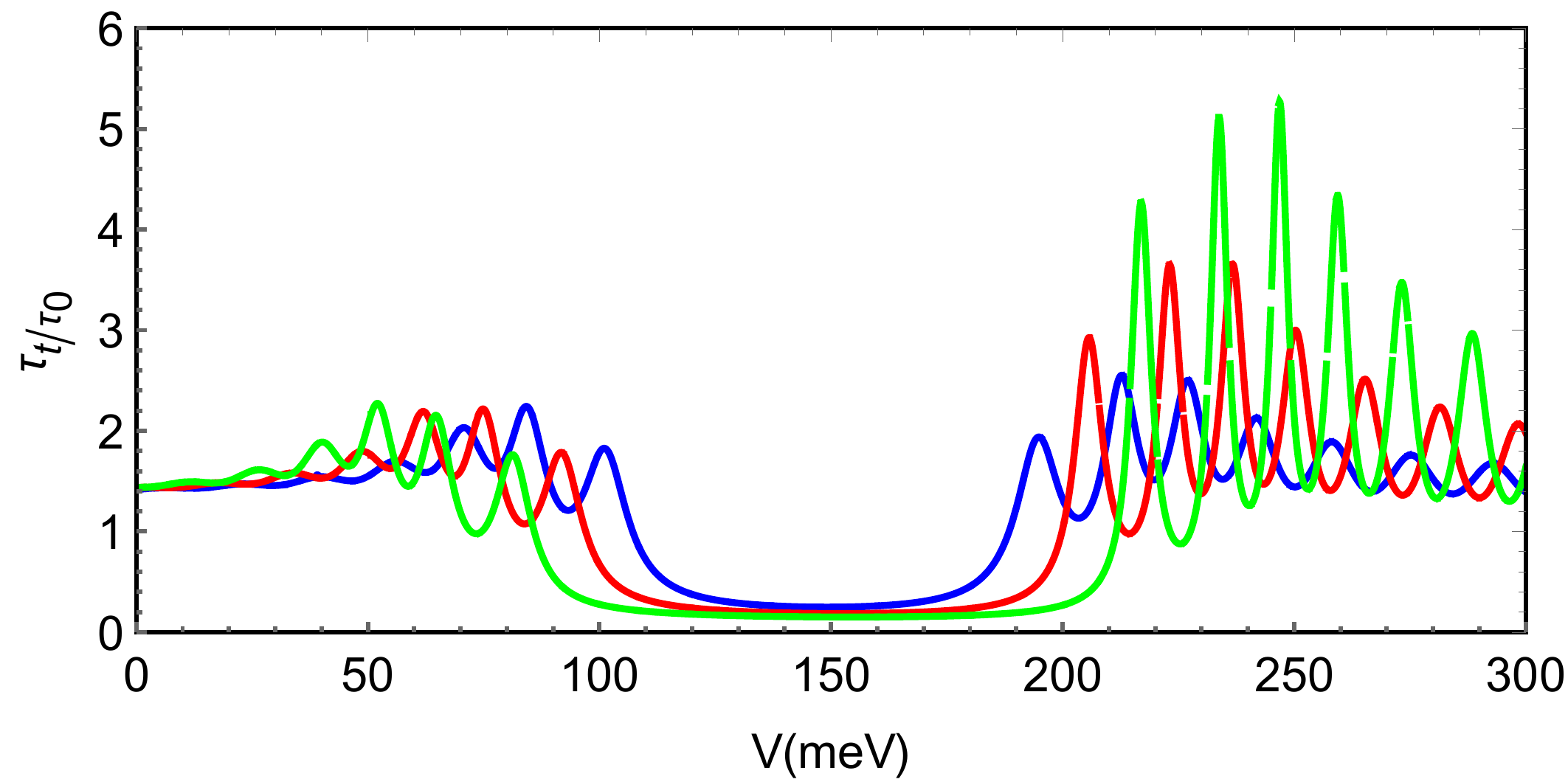}
		\label{fig3f}}
	\caption{{(color online) The group delay time   $\tau_{t}/\tau_{0}$ versus the barrier height $V$ for  $L=80$ nm, $\Delta=10$ meV and  $B=0.8$ T.  \textbf{\color{red}{(a,b)}}:
			$S=0$ (blue), $S=0.05$ (red),  $S=0.15$ (green)
			with  	$E=150$ meV, $\phi=10^{\circ}$.	 \textbf{\color{red}{(c,d)}}:
			$E=150$ meV (blue), $E=170$ meV (red),  $E=180$ meV (green) with   $S=0.1$, $\phi=10^{\circ}$.	
		\textbf{\color{red}{(e,f)}}:
			$\phi=5^{\circ}$ (blue), $\phi=10^{\circ}$, (red), $\phi=15^{\circ}$, (green) with $E=150$ meV,  $S=0.1$.	 \textbf{\color{red}{(a,c,e)}}: Armchair direction and 
			\textbf{\color{red}{(b,d,f)}}:
			zigzag direction. }}
	\label{fig3}
\end{figure}
The group delay time in transmission $\tau_{t}/\tau_{0}$ versus the barrier height $V$ is shown in Fig. \ref{fig3} for the armchair (\ref{fig3a},\ref{fig3c},\ref{fig3e}) and zigzag directions (\ref{fig3b},\ref{fig3d},\ref{fig3f}). 
In Figs. (\ref{fig3a},\ref{fig3b}), we see that increasing $V$ for different values of strain $S=0$ (blue), $S=0.05$ (red), and $S=0.15$ (green), with $E=150$ meV, $\Delta=10$ meV, $B=0.8$ T, $L=80  $ nm, and $\phi=10^{\circ}$ changes the group delay in transmission. 
Fig. (\ref{fig3c},\ref{fig3d}) depict the impact of incident energy $E$ on group delay in transmission versus $V$. The taken values are $E=150$ meV (blue), $E=170$ meV (red) and $E=180$ meV (green) with $\Delta=10$ meV, $B=0.8$ T, $L=80$ nm, $S=0.1$ and $\phi=10^{\circ}$. 
Figs. (\ref{fig3e},\ref{fig3f}) show that increasing $ V $ changes the group delay for three incident angle values: 
$\phi=5^{\circ}$ (blue), $\phi=10^{\circ}$ (red), $\phi=15^{\circ}$ (green) with $E=150$ meV, $\Delta=10$ meV, $B=0.8$ T, $L=80$ nm and $S=0.1$. 
When $V<50$ meV, $\tau_{t}/\tau_{0}$  is always equal to 1, electrons propagate through the barrier with Fermi velocity $v_F$ for the armchair directions (\ref{fig3b},\ref{fig3d},\ref{fig3f}), and a value greater than 1 for the zigzag  directions (\ref{fig3b},\ref{fig3d},\ref{fig3f}). 
In all figures, $\tau_{t}/\tau_{0} >1$  exhibits small oscillations, as for the interval 50 meV $<V<$ 100 meV. Then the wave function inside the barrier propagates rather than being evanescent. This shows that the group delay in transmission is subluminal.
When 100 meV $<V< $ 180 meV, the group delay time is $\tau_{t}/\tau_{0} <1$, indicating that electrons can travel faster than $\upsilon_F$, where $\upsilon_F$ is equivalent to the speed of light $c$ in optics. 
Furthermore, as  $V$ increases,  $\tau_{t}/\tau_{0}$ decreases exponentially, then rapidly approaches constant zero when $V=E$. 
When $ V $ becomes large enough, $\tau_{t}/\tau_{0}$ increases after oscillating with increasing $ V $, and the peak values become large compared to peaks in the range of 100 meV $ <V< $ 180 meV, particularly in the case of zigzag strain. 
We conclude that the value and number of peaks of $\tau_{t}/\tau_{0}$ increase as the values of $S$, $E$, and $\phi$ increase, particularly for strain along the zigzag direction.

   \begin{figure}[H]
   	\centering
   	\subfloat[$S=0$, $S=0.1$, $S=0.15$]{
   		\centering
   		\includegraphics[width=8cm, height=4.8cm]{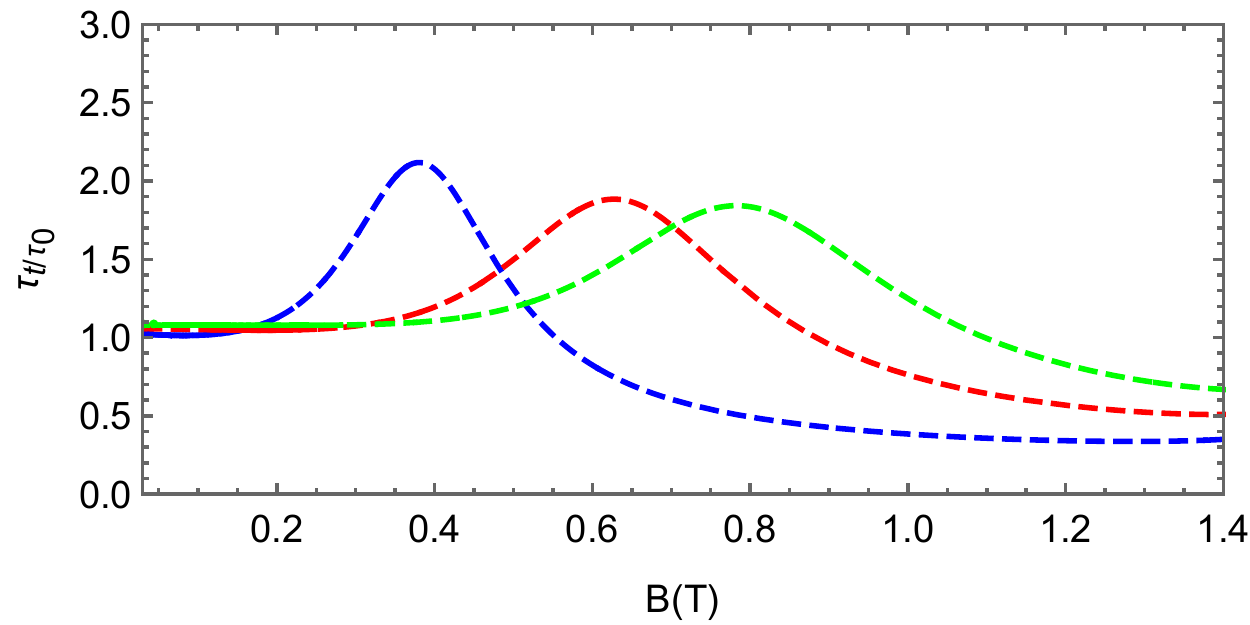}
   		\label{fig4a}
   	}\hspace{1cm}
   \subfloat[$S=0$, $S=0.1$, $S=0.15$]{
   		\centering
   		\includegraphics[width=8cm, height=4.8cm]{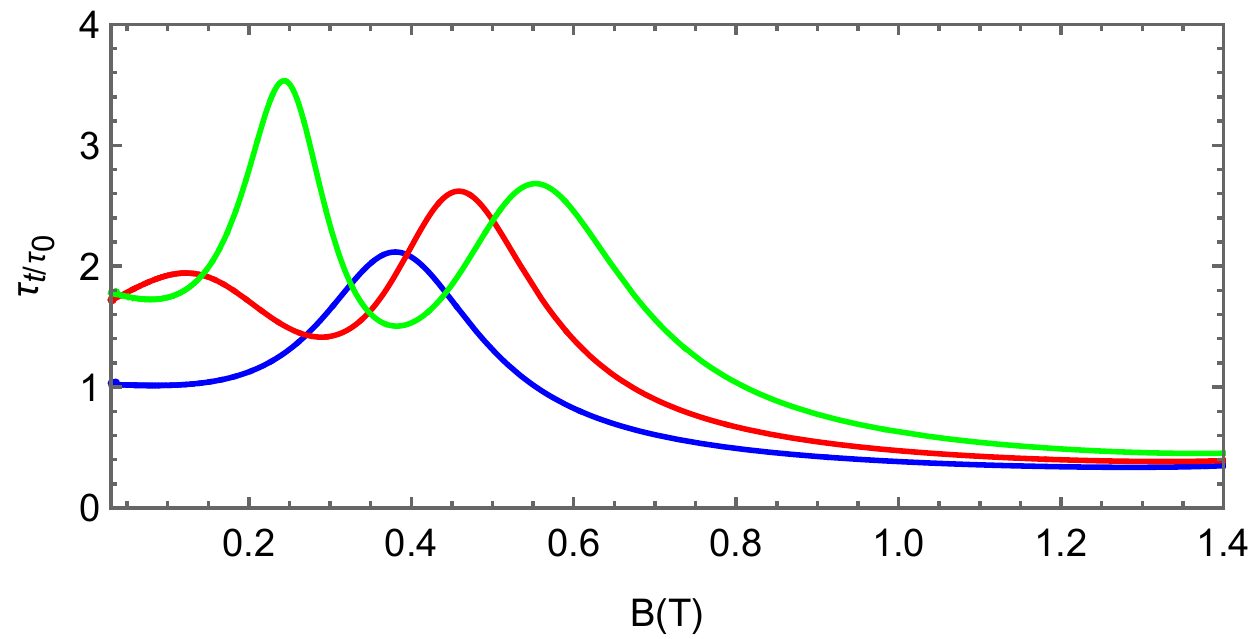}
   		\label{fig4b}}\\
   	\centering
   	\subfloat[$E=150$ meV, $E=170$ meV, $E=180$ meV]{
   		\centering
   		\includegraphics[width=8cm, height=4.8cm]{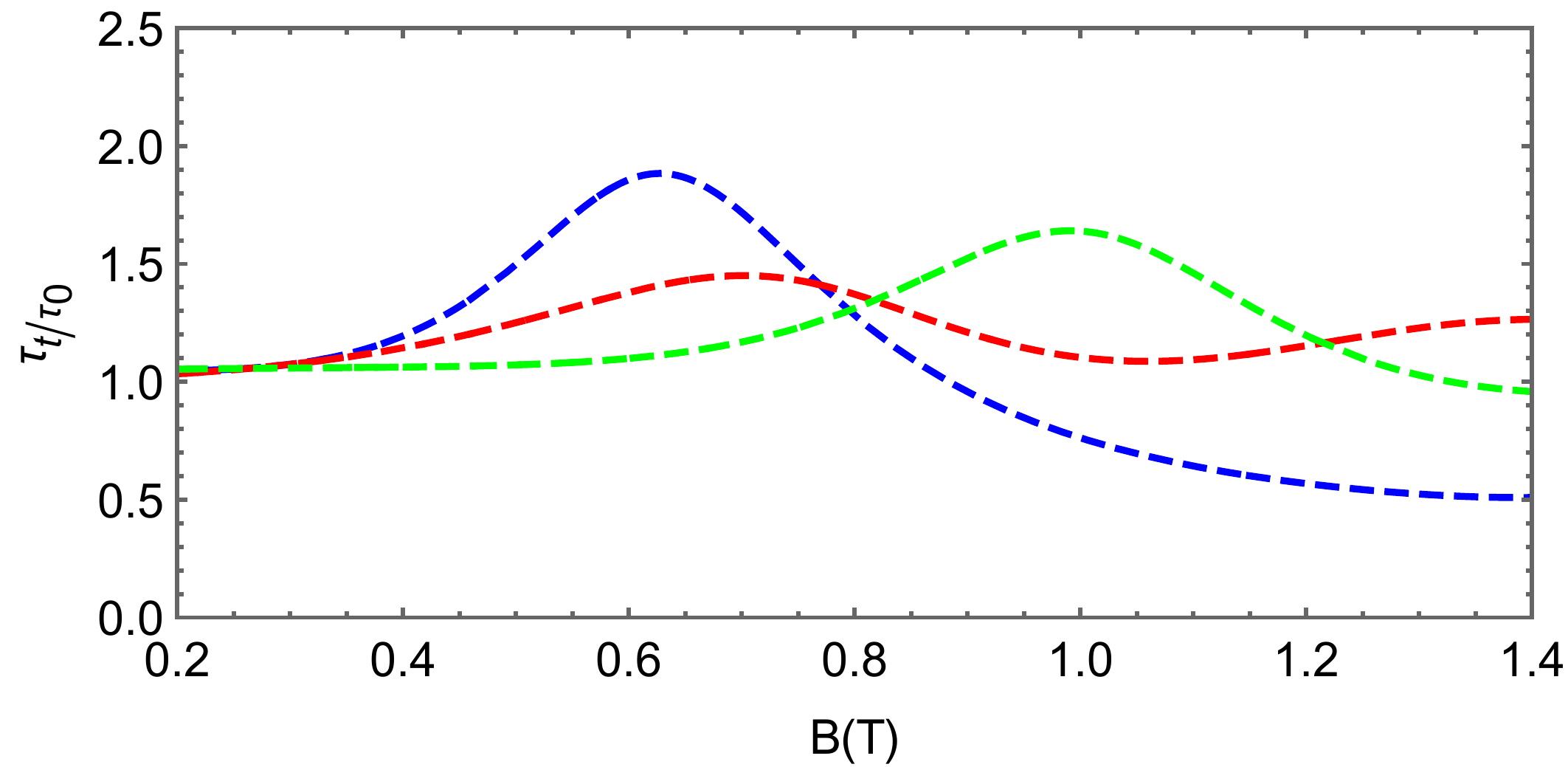}
   		\label{fig4c}
   	}\hspace{1cm}
   \subfloat[$E=150$ meV, $E=170$ meV, $E=180$ meV]{
   		\centering
   		\includegraphics[width=8cm, height=4.8cm]{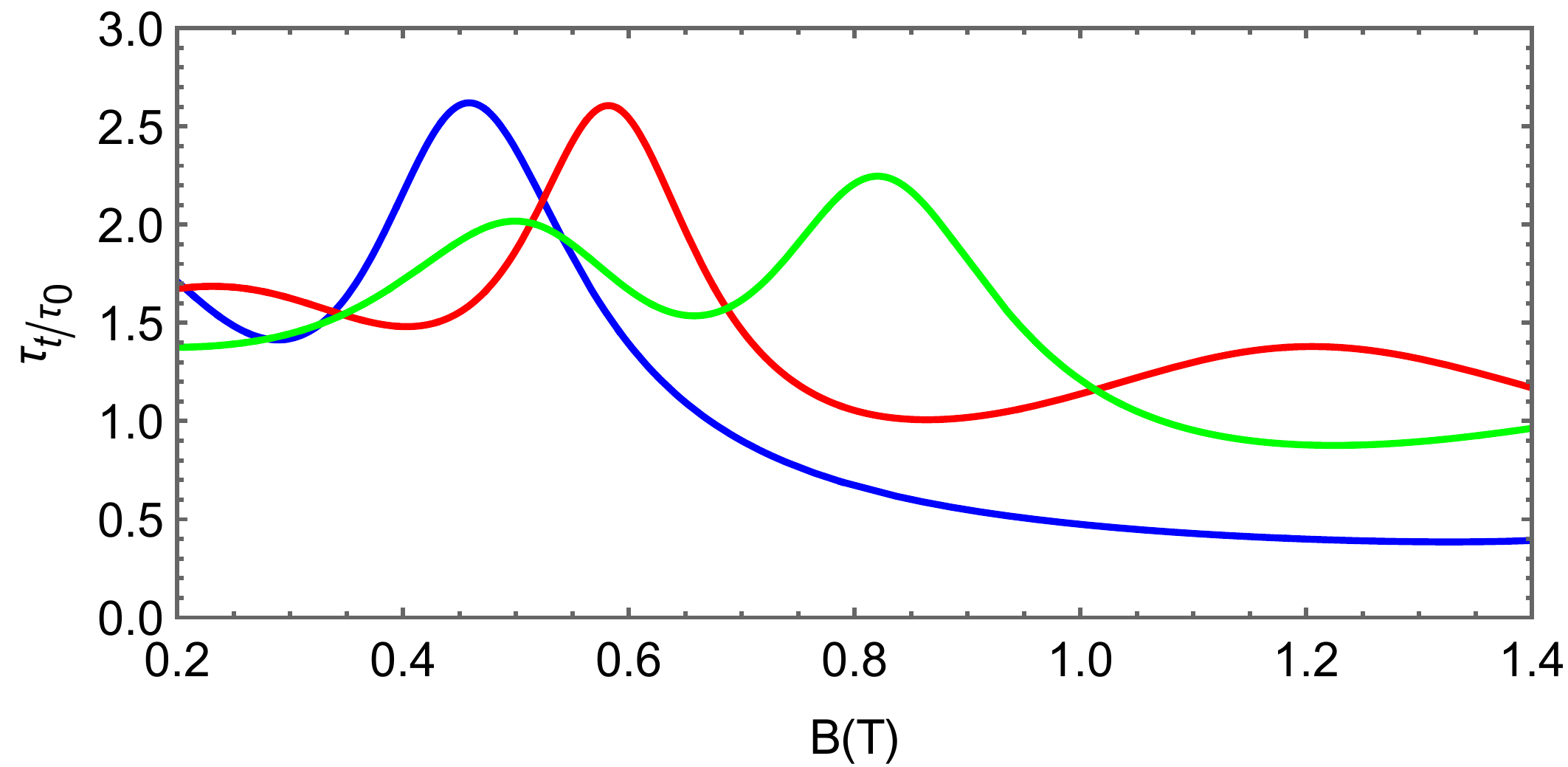}
   		\label{fig4d}}\\ 
   	\centering
   	\subfloat[$\phi=0^{\circ}$, $\phi=5^{\circ}$, $\phi=10^{\circ}$]{
   		\centering
   		\includegraphics[width=8cm, height=4.8cm]{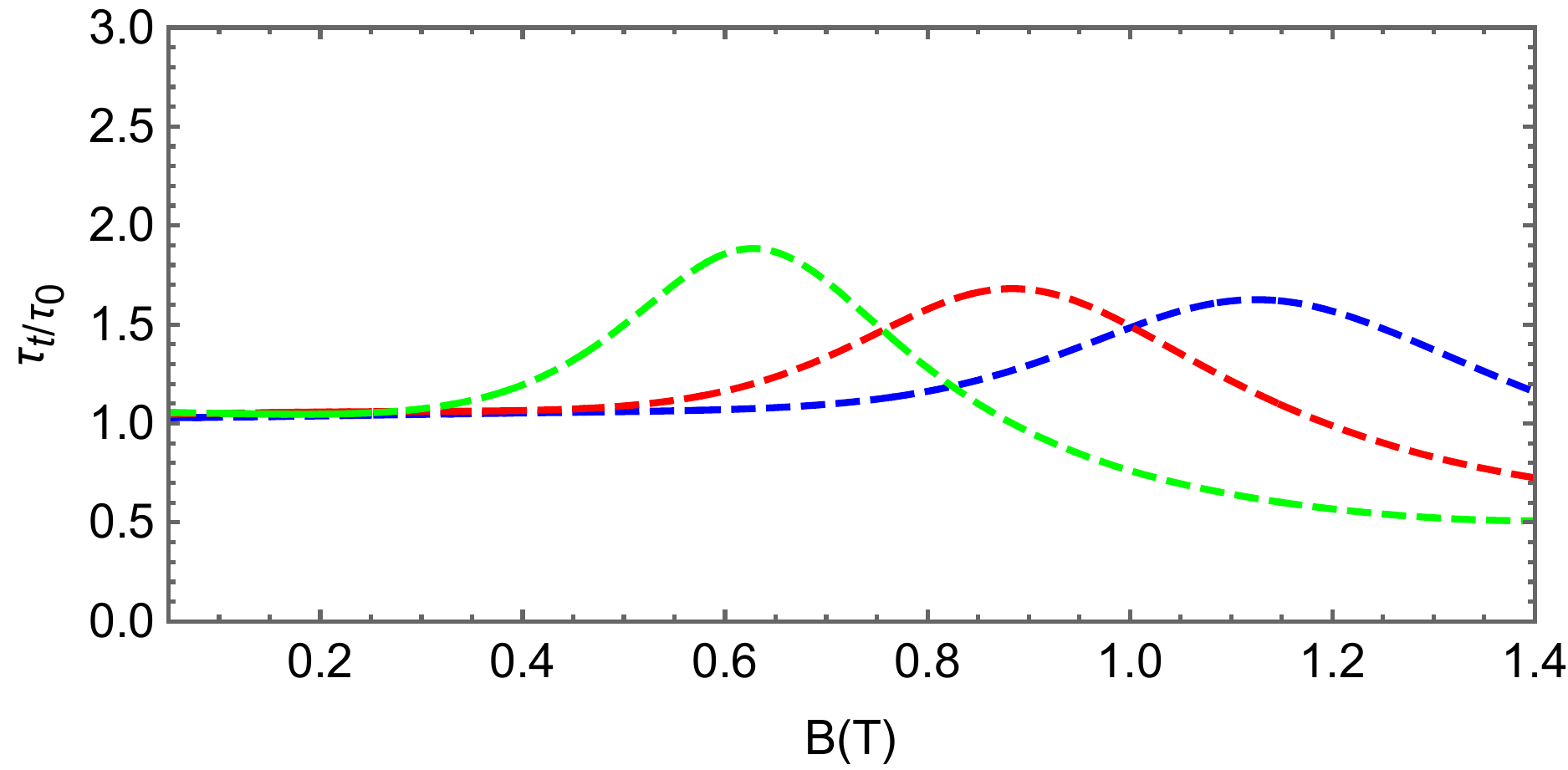}
   		\label{fig4e}
   	}\hspace{1cm}
   \subfloat[$\phi=0^{\circ}$, $\phi=5^{\circ}$, $\phi=10^{\circ}$]{
   		\centering
   		\includegraphics[width=8cm, height=4.8cm]{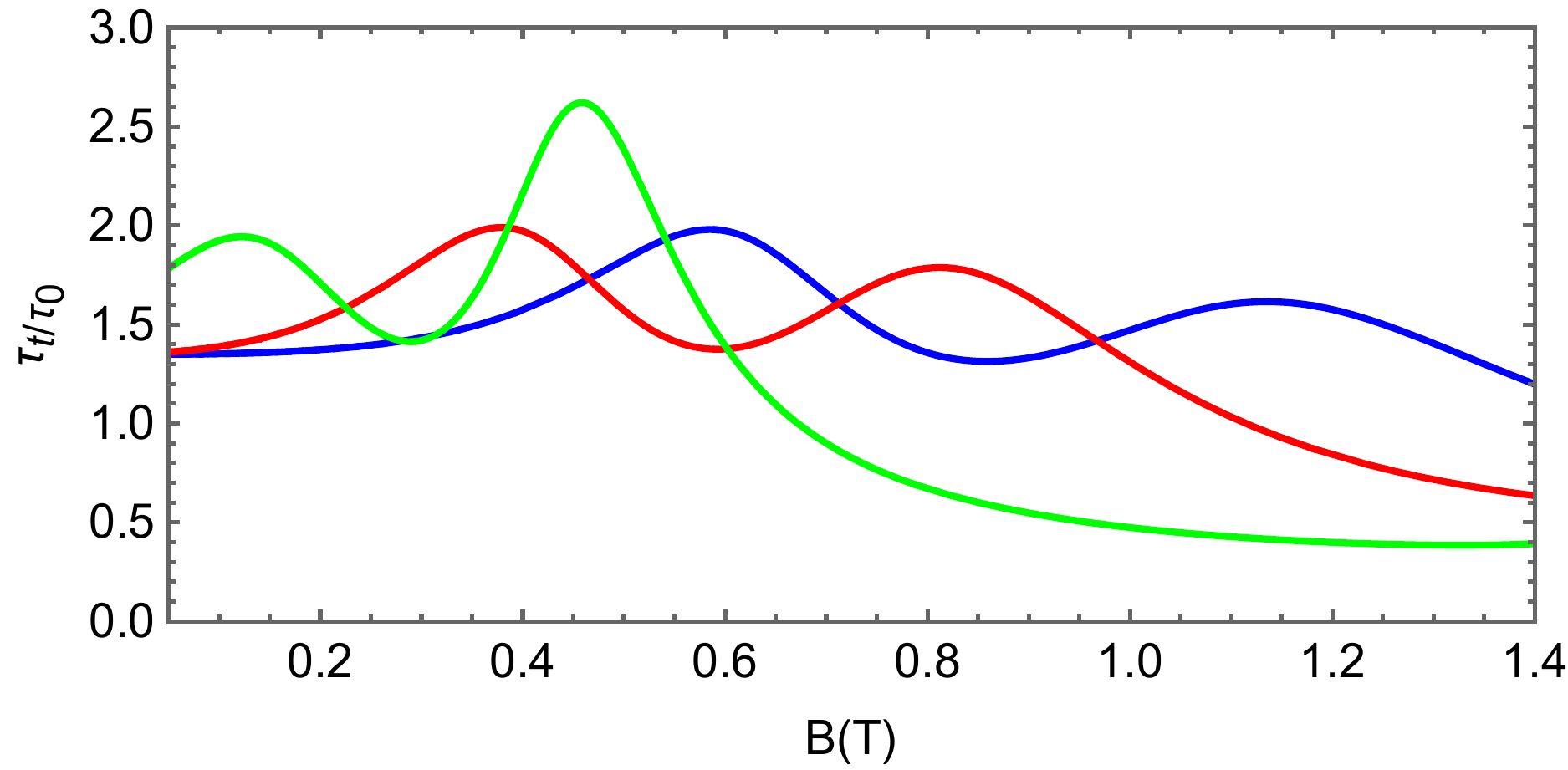}
   		\label{fig4f}}
   	\caption{{(color online) The group delay time in transmission $\tau_{t}/\tau_{0}$ versus the magnetic field  $B$ with $V=100$ meV,  $L=80$ nm and $\Delta=10$  meV.  \textbf{\color{red}{(a,b)}}:
   			 $S=0$ (blue), $S=0.1$, (red),  $S=0.15$, (green) with $E=150$ meV, $\phi=10^{\circ}$.  \textbf{\color{red}{(c,d)}}: $E=150$ meV (blue), $E=170$  meV, (red),  $E=180$ meV, (green) with  $S=0.1$, $\phi=10^{\circ}$.  \textbf{\color{red}{(e,f)}}: $\phi=0^{\circ}$ (blue), $\phi=5^{\circ}$ (red), $\phi=10^{\circ}$ (green) with $E=150$ meV, $S=0.1$.  \textbf{\color{red}{(a,c,e)}}: 	Armchair strain direction and 
   			\textbf{\color{red}{(b,d,f)}}:
   			zigzag strain direction. }}
   	\label{fig4}
   \end{figure}
The group delay time in transmission $\tau_{t}/\tau_{0}$ versus the magnetic field $B$ for the armchair (\ref{fig4a},\ref{fig4c},\ref{fig4e}) and zigzag directions (\ref{fig4b},\ref{fig4d},\ref{fig4f}) is depicted in Fig \ref{fig4}. 
Figs. (\ref{fig4a},\ref{fig4b}) depict the effect of $B$ on $\tau_{t}/\tau_{0}$ for various strain $S=0$ (blue), $S=0.1$ (red), and $S=0.15$ (green), with $V=100$ meV, $E=150$ meV, $\Delta=10$ meV, and $\phi=10^{\circ}$.    
Figs. (\ref{fig4c},\ref{fig4d}) depict the effect of different incident energy values on $\tau_{t}/\tau_{0}$  versus $B$, with $\Delta=10$ meV, $V=100$ meV, $L=80$ nm, $S=0.1$, and $\phi=10^{\circ}$. 
The effect of the incident angle 	$\phi=5^{\circ}$ (blue), $\phi=10^{\circ}$ (red), $\phi=15^{\circ}$ (green) is illustrated in Figs. (\ref{fig4e},\ref{fig4f}) where $E=150$~meV, $\Delta=10$ meV, $V=100$ meV, $L=80$ nm and $S=0.1$.
As shown in Figs. (\ref{fig4a},\ref{fig4b},\ref{fig4c},\ref{fig4e}), $\tau_{t}/\tau_{0}$ decreases exponentially as long as $B$ increase. 
As $ B $ increases more, $\tau_{t}/\tau_{0}$ becomes independent of the magnetic field, which is a manifestation of  Hartman effect. 
We see that the works \cite{YueBan, Fattasse} have both reported on similar behavior. In contrast, $\tau_{t}/\tau_{0}$ presents oscillatory behavior against $B$ in the zigzag direction, as shown in Figs. (\ref{fig4b},\ref{fig4d},\ref{fig4f}).
This is due to the fact that the transmitted wave, which is a hybrid of the oscillating and evanescent waves, will alter $\tau_{t}/\tau_{0}$ in the barrier zone. As a result,  $\tau_{t}/\tau_{0}$ is affected by both waves. 
It is important to keep in mind that oscillating waves produce resonant peaks. An exponential decline is, however, produced by the evanescent waves.  Another observation is that the number of peaks and oscillations increases with increasing of $B$, for the zigzag directions compared with armchair directions

\begin{figure}[H]
	\centering
	\subfloat[$S=0.05$, $S=0.1$, $S=0.15$]{
		\centering
		\includegraphics[width=7cm, height=4.6cm]{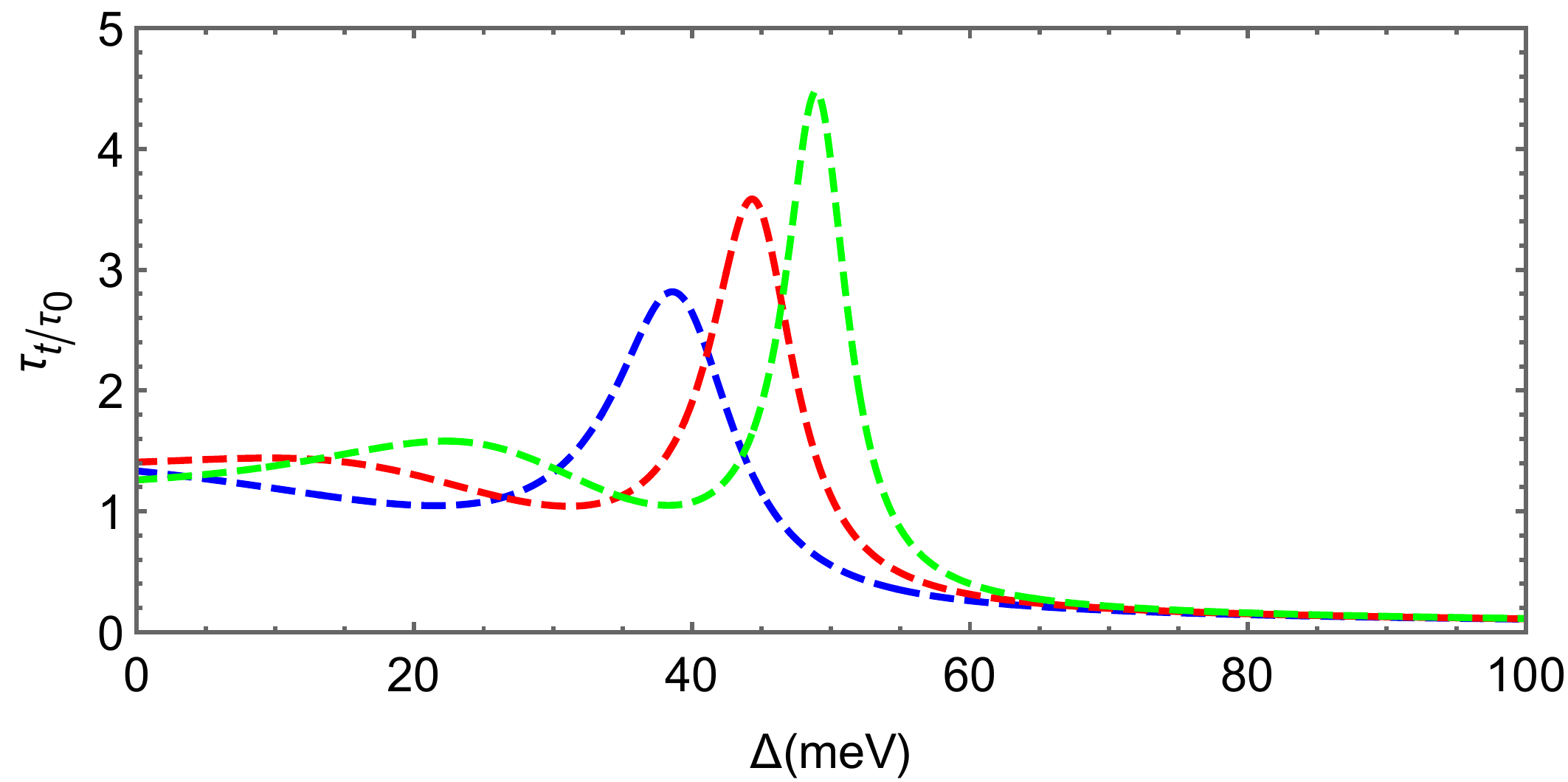}
		\label{fig5a}
	}\hspace{1cm}
\subfloat[$S=0.05$, $S=0.1$, $S=0.15$]{
		\centering
		\includegraphics[width=7cm, height=4.6cm]{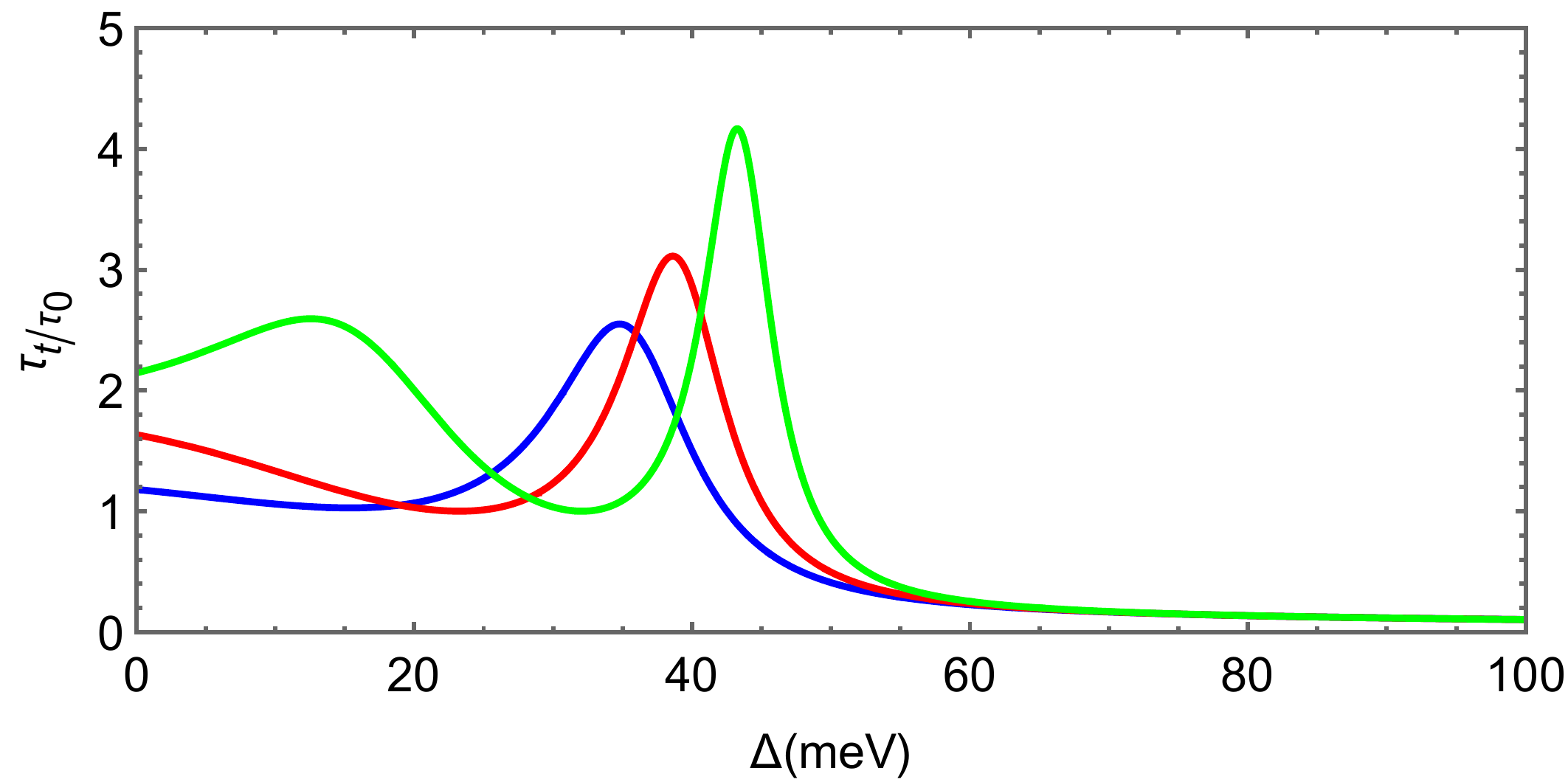}
		\label{fig5b}}\\
	\centering
	\subfloat[$E=150$ meV, $E=170$ meV, $E=180$ meV]{
		\centering
		\includegraphics[width=7cm, height=4.6cm]{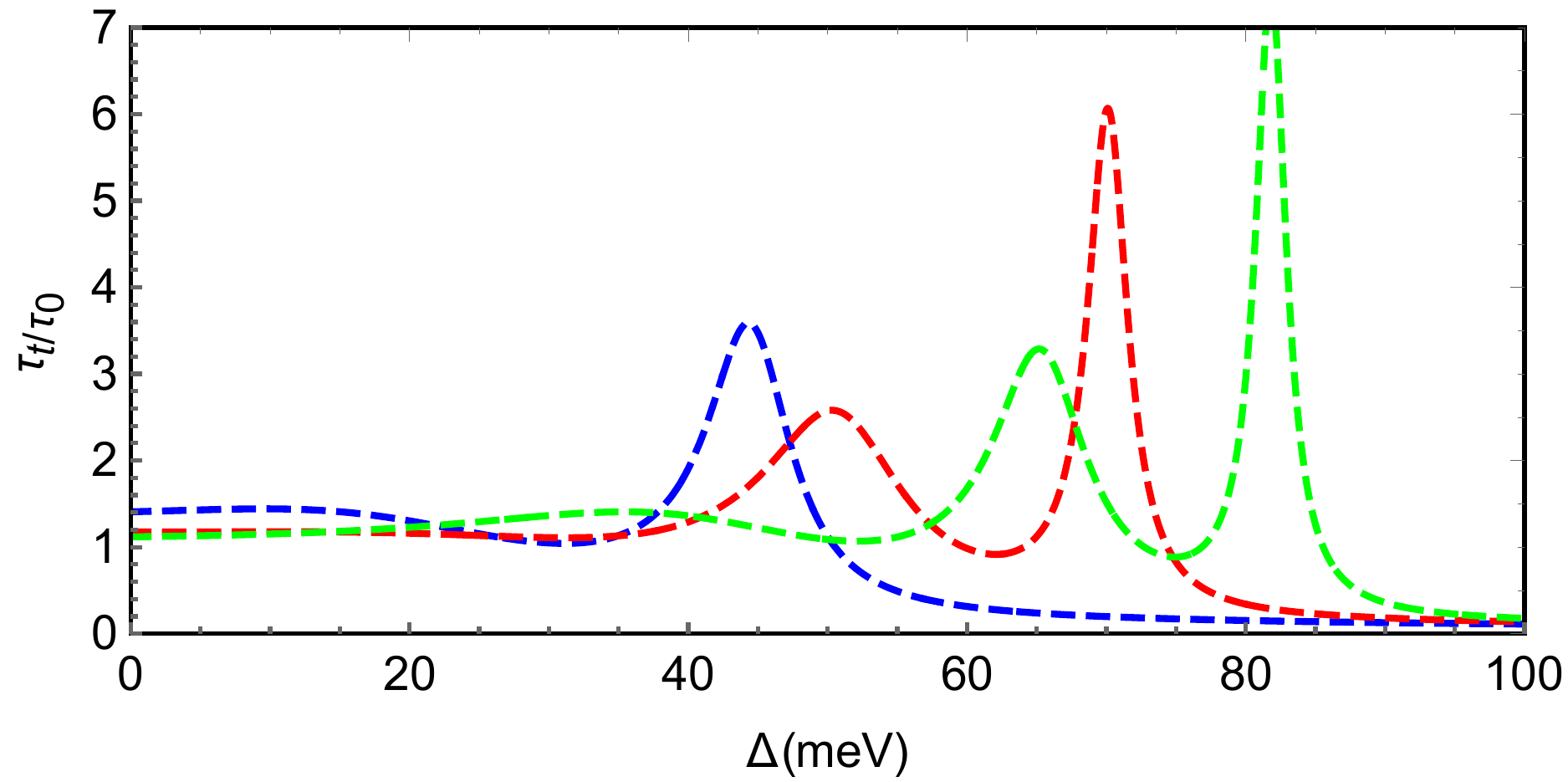}
		\label{fig5c}
	}\hspace{1cm}
\subfloat[$E=150$ meV, $E=170$ meV, $E=180$ meV]{
		\centering
		\includegraphics[width=7cm, height=4.6cm]{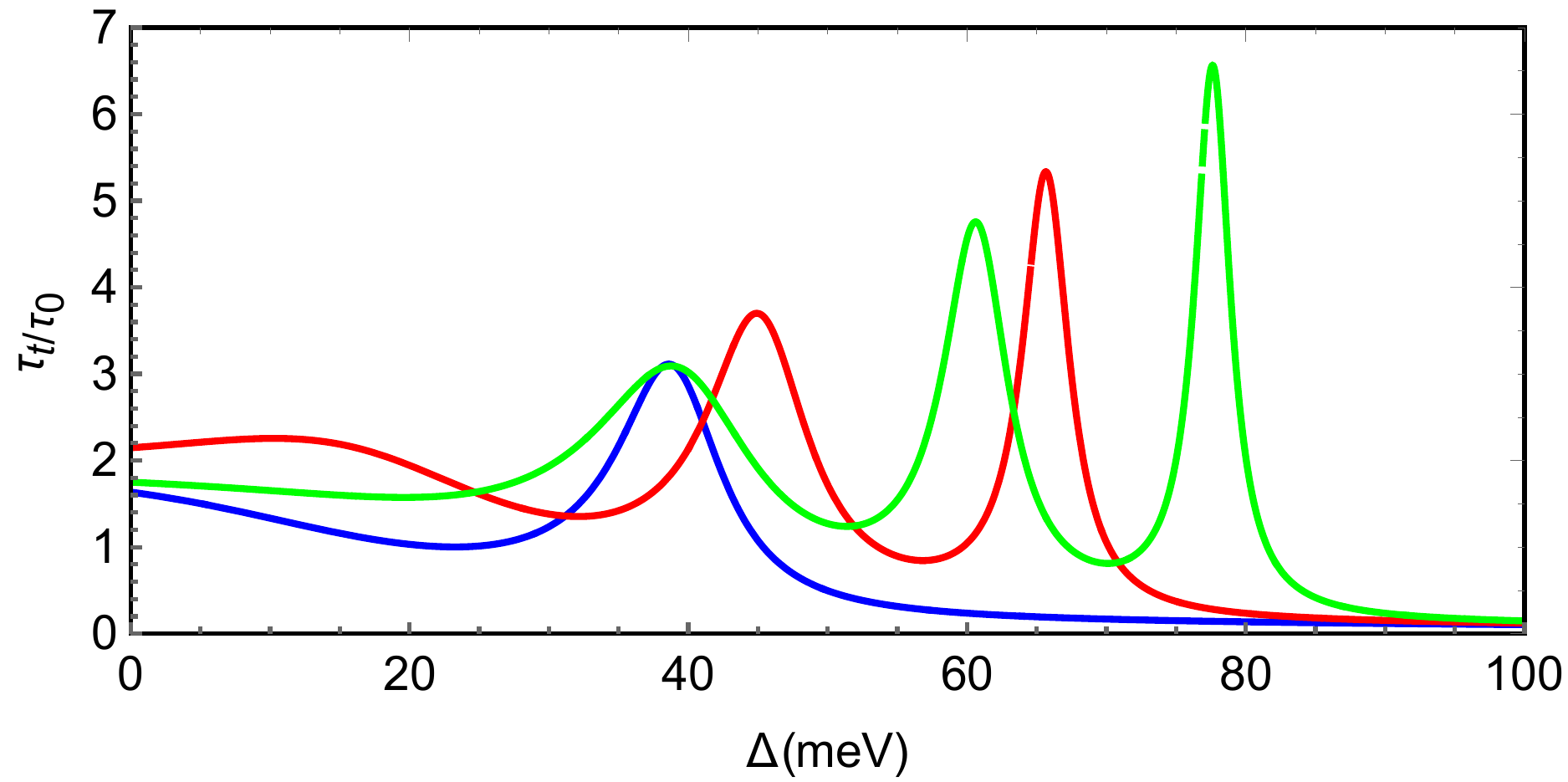}
		\label{fig5d}}\\
	\centering
	\subfloat[$\phi=0^{\circ}$, $\phi=5^{\circ}$, $\phi=10^{\circ}$]{
		\centering
		\includegraphics[width=7cm, height=4.6cm]{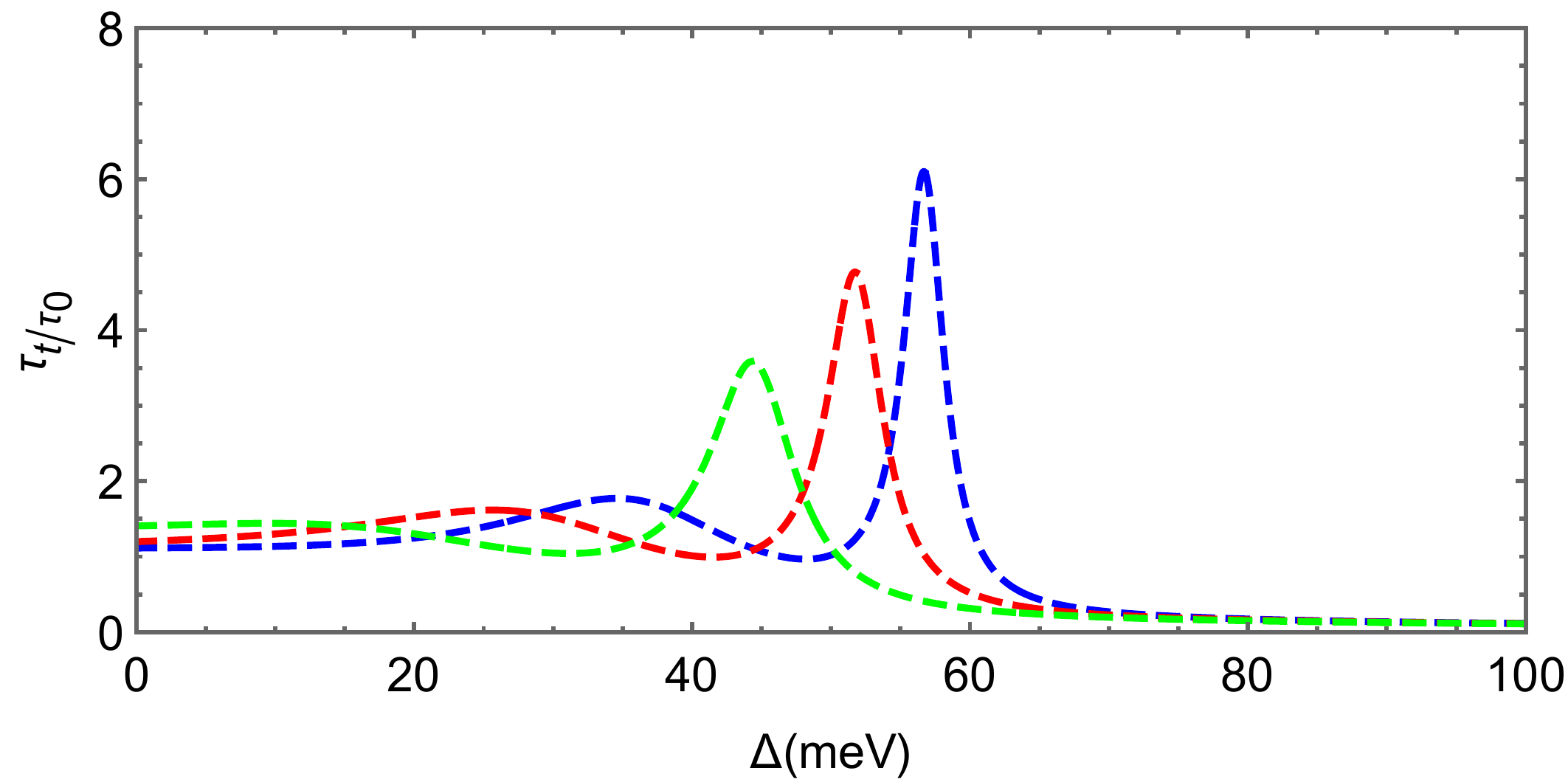}
		\label{fig5e}
	}\hspace{1cm}
\subfloat[$\phi=0^{\circ}$, $\phi=5^{\circ}$, $\phi=10^{\circ}$]{
		\centering
		\includegraphics[width=7cm, height=4.6cm]{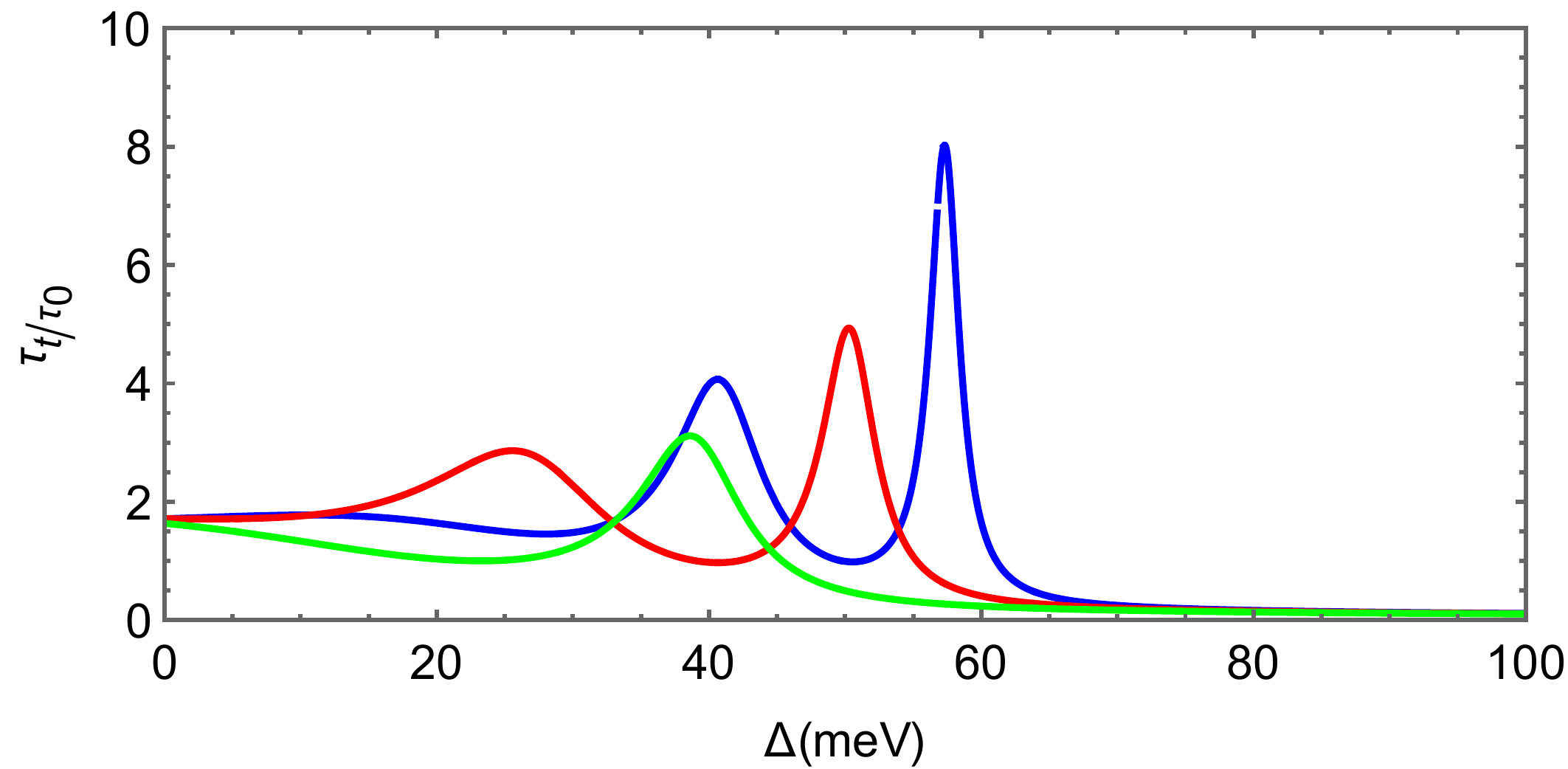}
		\label{fig5f}}
	\caption{{(color online) The group delay time  $\tau_{t}/\tau_{0}$ versus the
			energy gap $\Delta$ for $V=80$ meV,  $L=80$ nm and  $B=0.8$ T.   \textbf{\color{red}{(a,b)}}:
			$S=0.05$ (blue), $S=0.1$ (red),  $S=0.15$ (green)
			with $E=150$ meV,   $\phi=10^{\circ}$. \textbf{\color{red}{(c,d)}}:
			$E=150$ meV (blue), $E=170$ meV (red),  $E=180$ meV (green) with   $S=0.1$, $\phi=10^{\circ}$.	
		\textbf{\color{red}{(e,f)}}: 
			$\phi=0^{\circ}$ (blue), $\phi=5^{\circ}$ (red), $\phi=10^{\circ}$ (green)
			with $E=150$ meV,  $S=0.1$.  \textbf{\color{red}{(a,c,e)}}: 	Armchair direction and
			\textbf{\color{red}{(b,d,f)}}:
			zigzag direction.  }}
	\label{fig5}
\end{figure}
Fig. \ref{fig5} plots the group delay time $\tau_{t}/\tau_{0}$ against the energy gap $\Delta$ using the same parameters as in Fig. \ref{fig3}. 
The particles propagate through the barrier with the Fermi velocity $v_F$ ($\tau_{t}/\tau_{0}= 1$) for a strain applied along the armchair direction for values of $\Delta \leq 30$ meV. 
However, as $\Delta$ rises and the wave propagates, $\tau_{t}/\tau_{0}$ exhibits an oscillating characteristic. This has to do with the self-interference delay, which is a consequence of the incident and reflected waves overlapping at the barrier's front region. When the $\Delta$ is large enough, $\tau_{t}/\tau_{0}$ decays exponentially to zero, regardless of the values of $S$, $ E $, and $\phi$. 
The same thing we see for the zigzag directions, except that the particles cross the barrier at a speed lower than the Fermi velocity $v_F$. Another noteworthy point is that $\tau_{t}/\tau_{0}$ has the highest value for normal incidence ($\phi=0^{\circ}$) when compared to the other $ \phi $ values.

\begin{figure}[H]
	\centering
	\subfloat[$S=0$, $S=0.1$, $S=0.15$]{
		\centering
		\includegraphics[width=7cm, height=4.8cm]{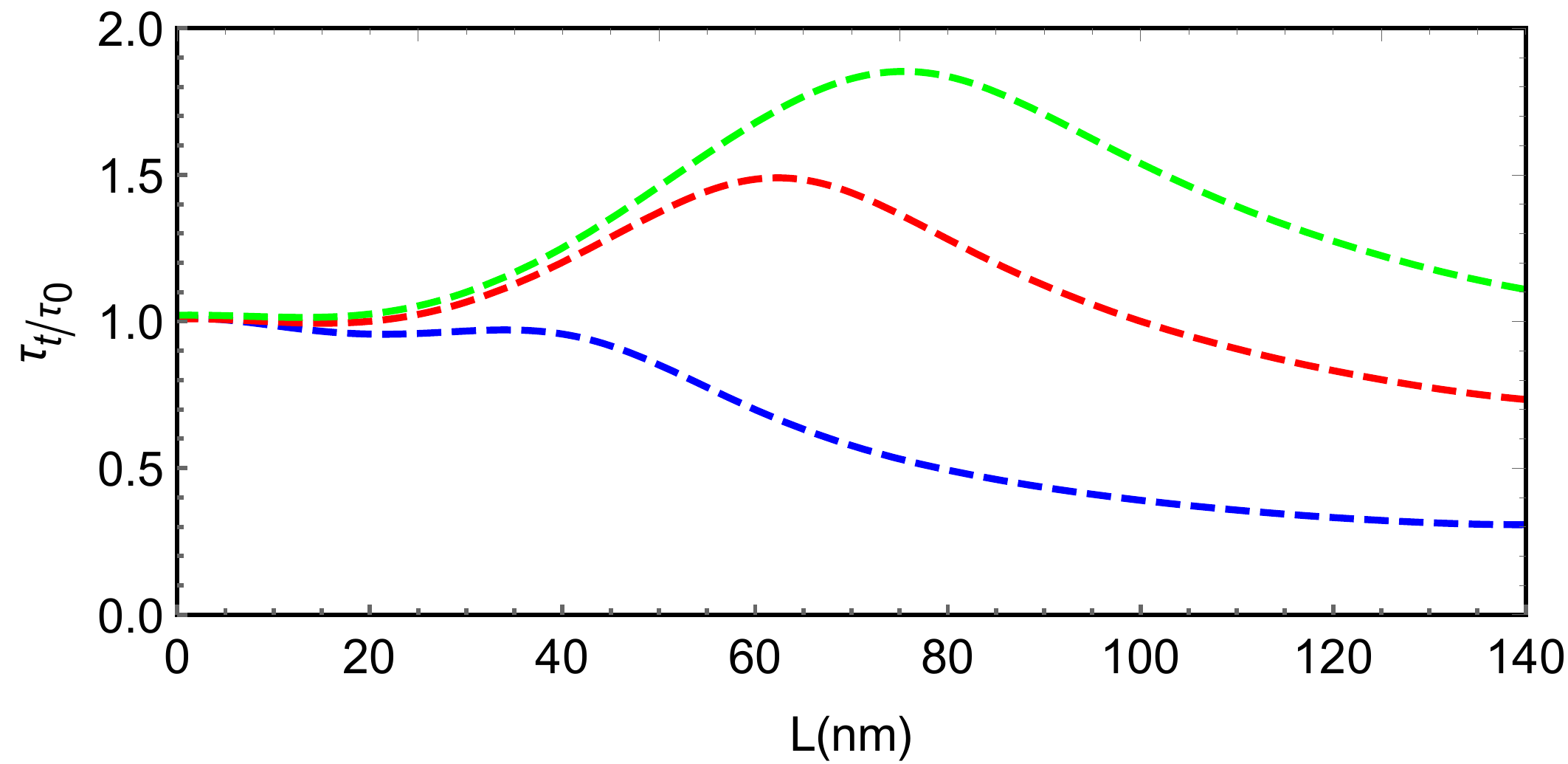}
		\label{fig6a}
	}\hspace{1cm}
\subfloat[$S=0$, $S=0.1$, $S=0.15$]{
		\centering
		\includegraphics[width=7cm, height=4.8cm]{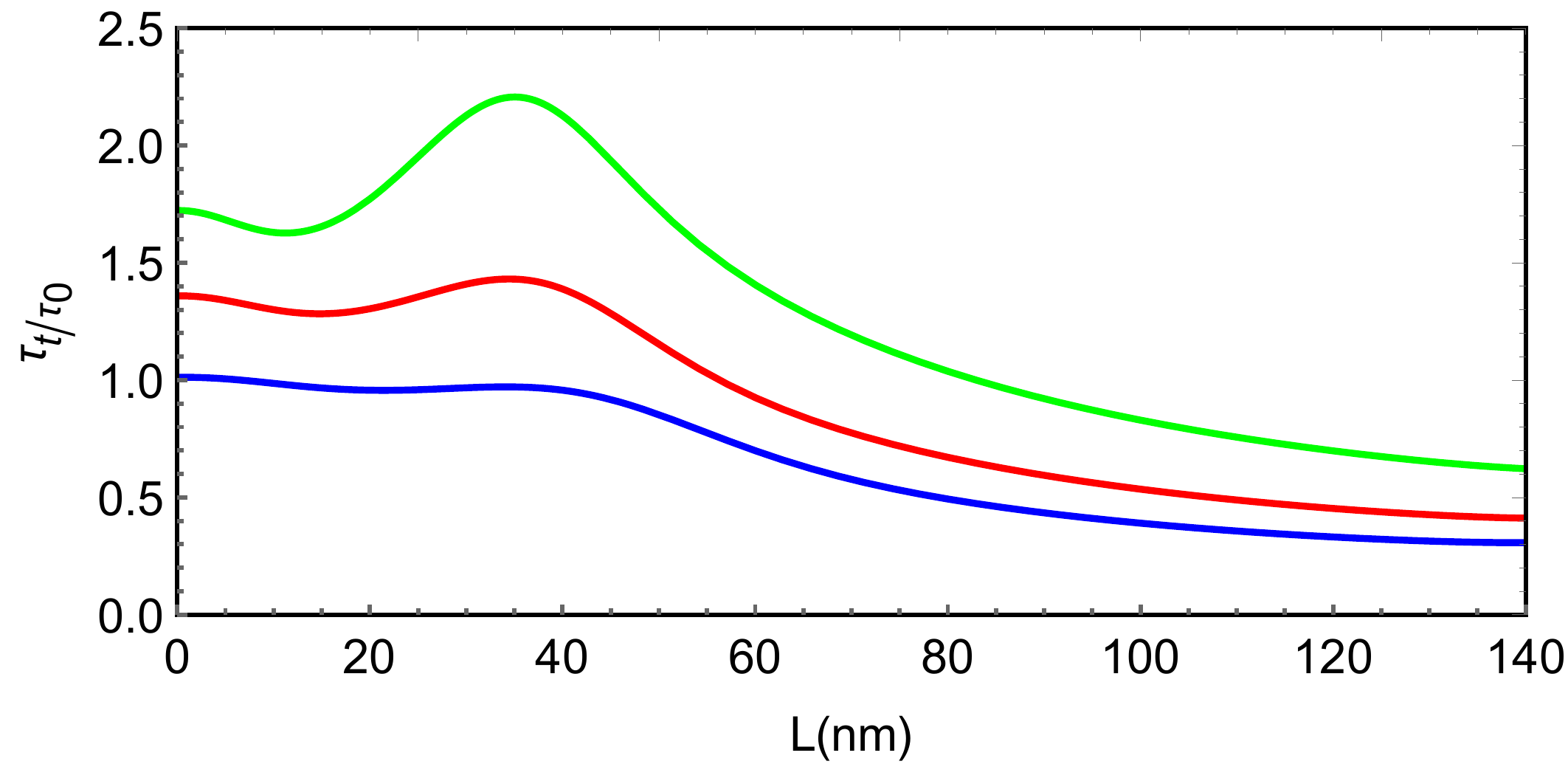}
		\label{fig6b}}\\
	\centering
	\subfloat[$E=150$ meV, $E=170$ meV, $E=180$ meV]{
		\centering
		\includegraphics[width=7cm, height=4.8cm]{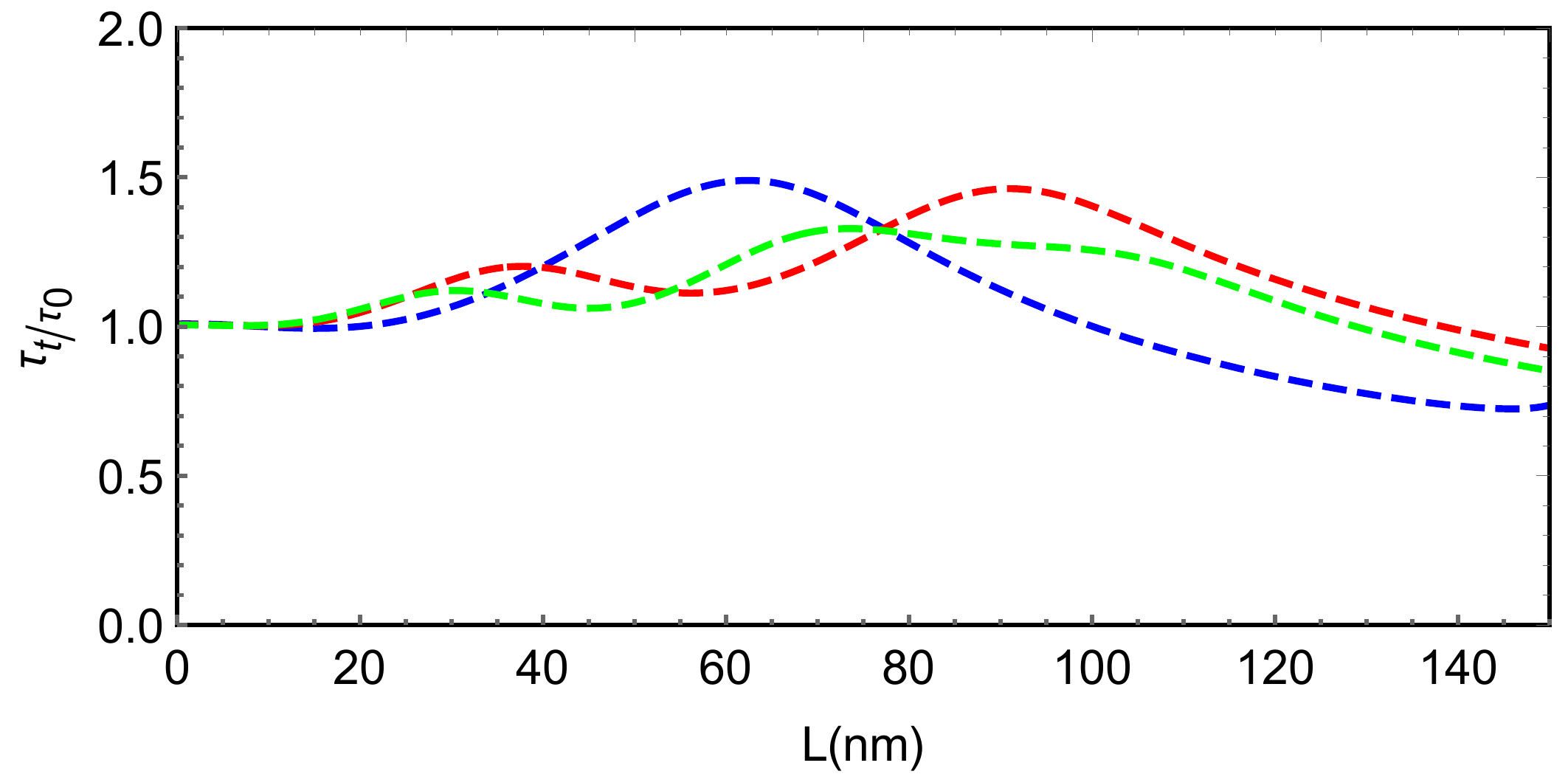}
		\label{fig6c}
	}\hspace{1cm}
\subfloat[$E=150$ meV, $E=170$ meV, $E=180$ meV]{
		\centering
		\includegraphics[width=7cm, height=4.8cm]{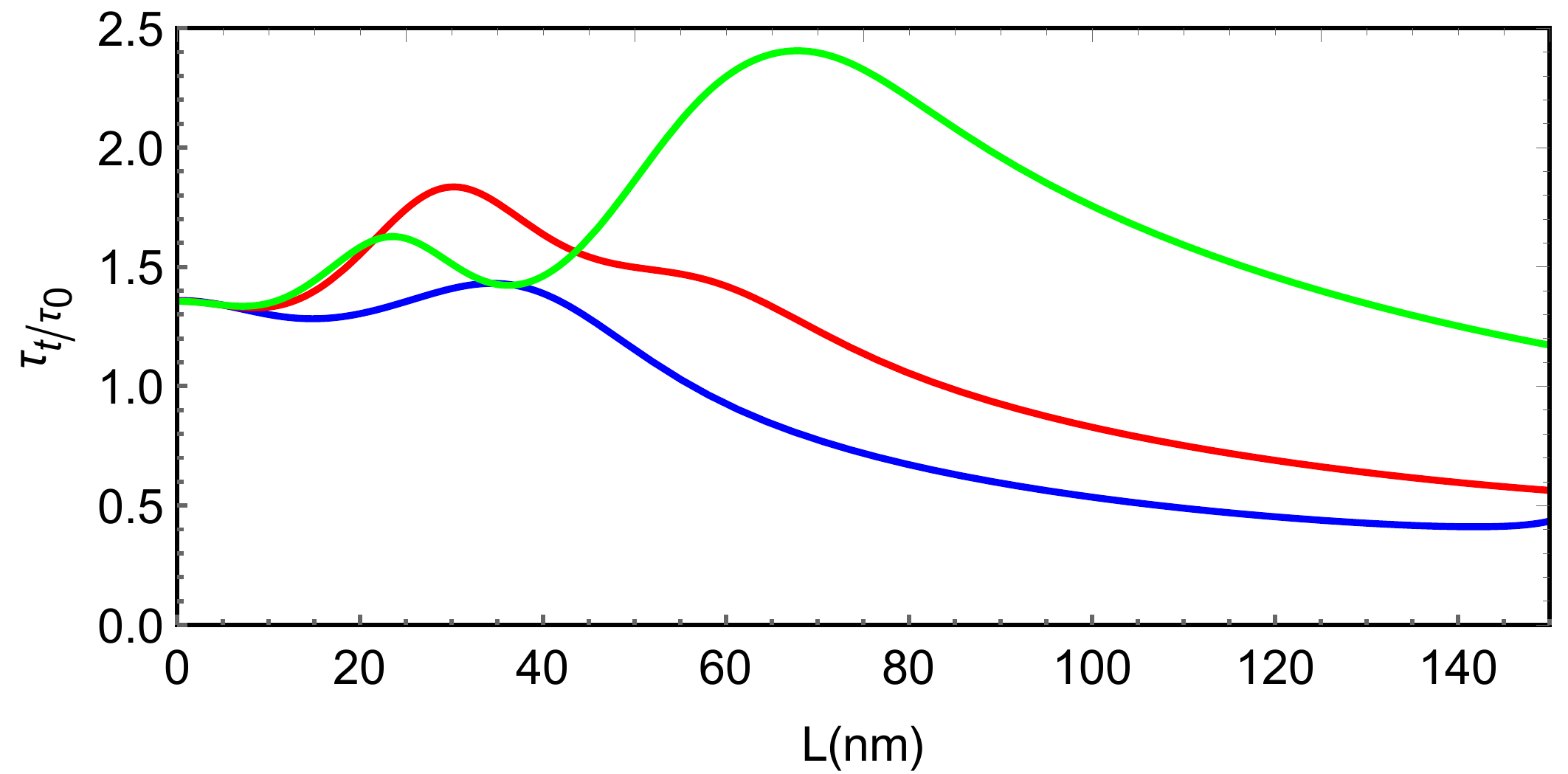}
		\label{fig6d}}\\
	\centering
	\subfloat[$\phi=0^{\circ}$, $\phi=5^{\circ}$, $\phi=10^{\circ}$]{
		\centering
		\includegraphics[width=7cm, height=4.8cm]{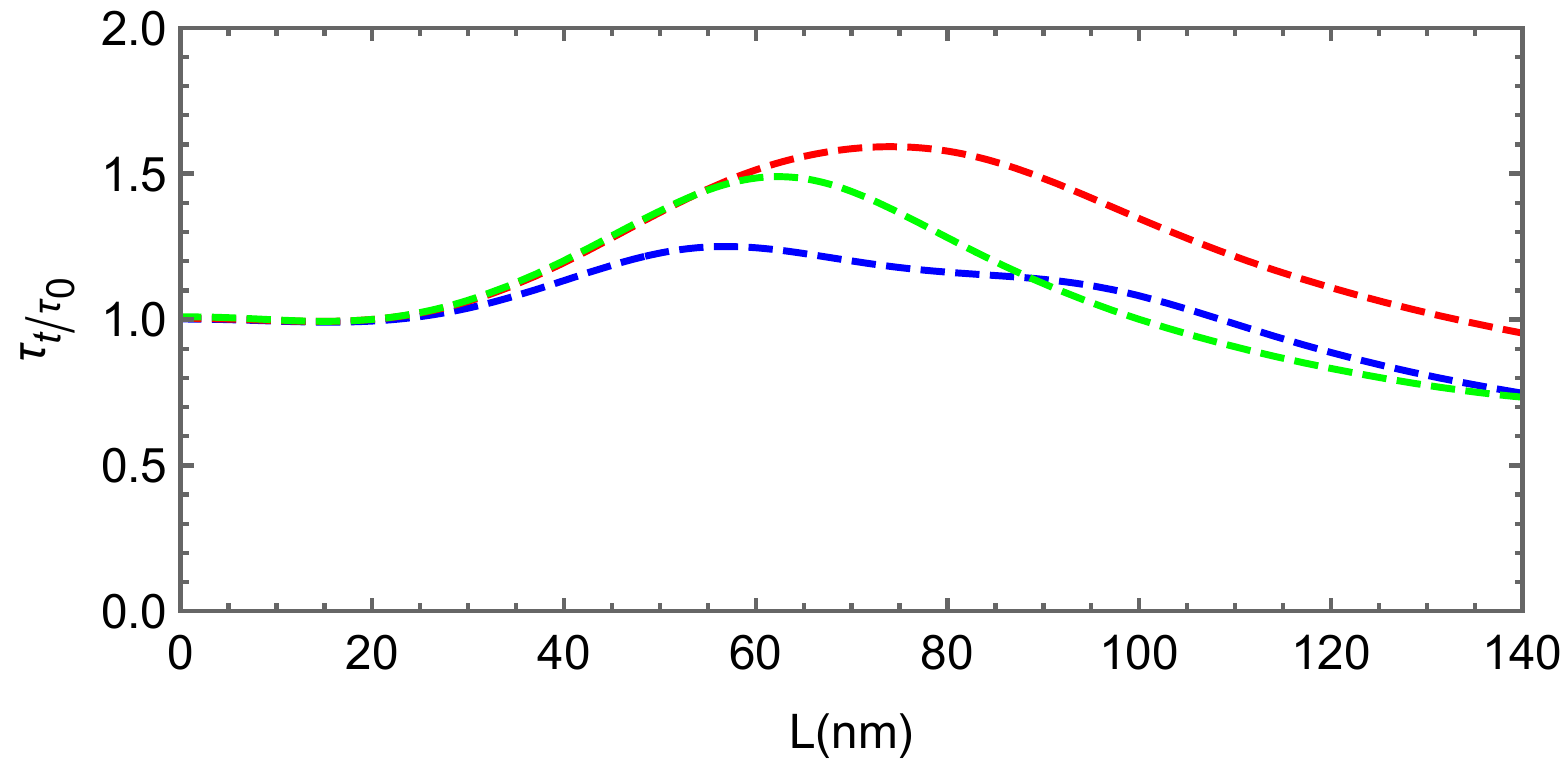}
		\label{fig6e}
	}
\hspace{1cm}
\subfloat[$\phi=0^{\circ}$, $\phi=5^{\circ}$, $\phi=10^{\circ}$]{
		\centering
		\includegraphics[width=7cm, height=4.8cm]{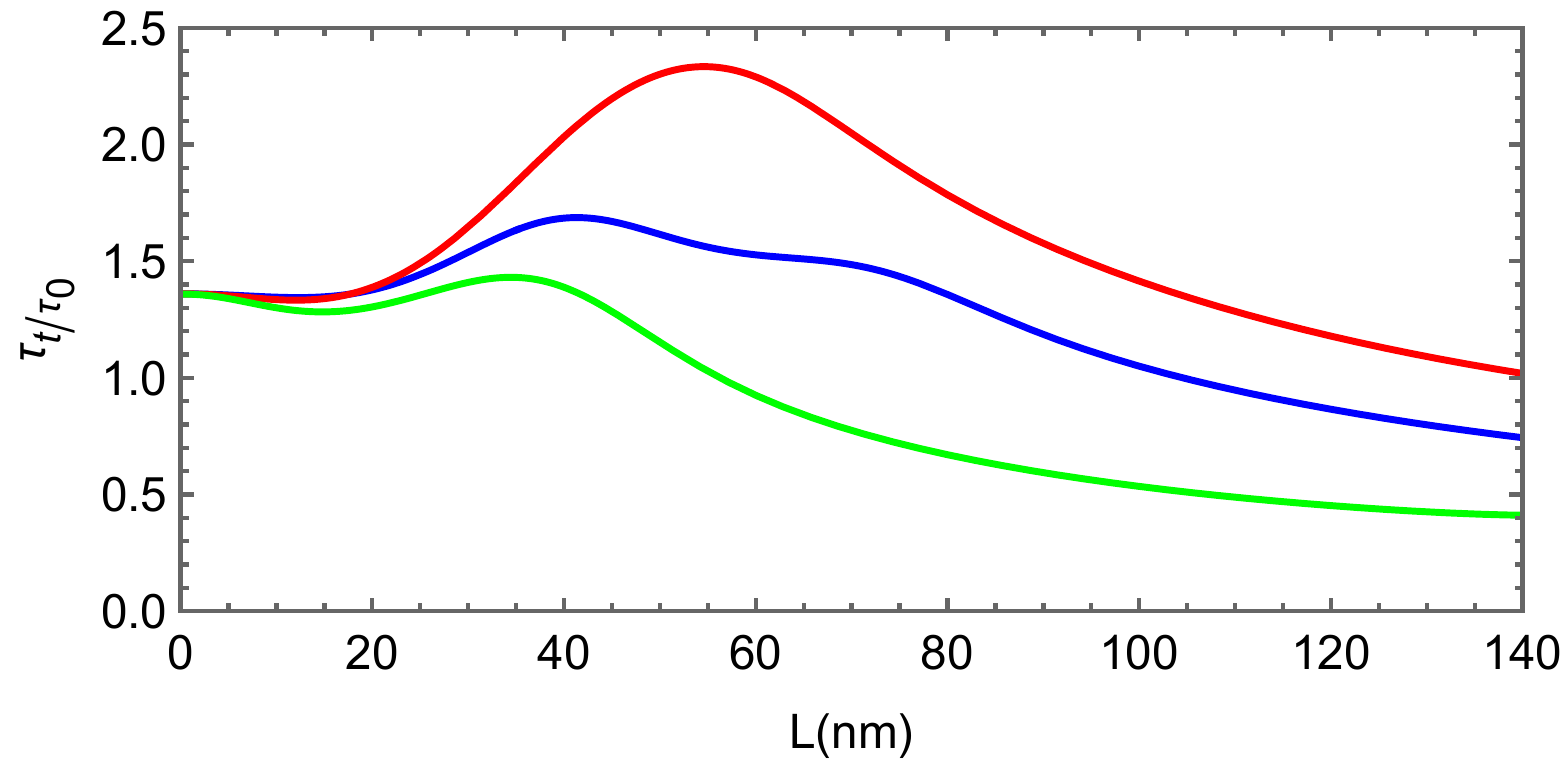}
		\label{fig6f}}
	\caption{{(color online) The group delay time  $\tau_{t}/\tau_{0}$ versus
			the barrier width $L$ for $V=100$ meV,  $B=0.8$ T and  $\Delta=10$ meV.   \textbf{\color{red}{(a,b)}}:
			$S=0$ (blue), $S=0.1$ (red),  $S=0.15$ (green)
			with $E=150$ meV, $\phi=10^{\circ}$.
			\textbf{\color{red}{(c,d)}}:
			 $E=150$ meV (blue), $E=170$ meV (red),  $E=180$ meV (green) with   $S=0.1$, $\phi=10^{\circ}$.	
			\textbf{\color{red}{(e,f)}}: 
			$\phi=0^{\circ}$ (blue), $\phi=5^{\circ}$ (red), $\phi=10^{\circ}$ (green)
			with $E=150$ meV, $S=0.1$.  \textbf{\color{red}{(a,c,e)}}: 	Armchair  direction and
			\textbf{\color{red}{(b,d,f)}}:
			zigzag  direction.  }}
	\label{fig6}
\end{figure}
The group delay time $\tau_{t}/\tau_{0}$ versus the barrier width $L$ is shown in Fig. \ref{fig6} using the same parameters as in Fig. \ref{fig3} except for the third incident energy value, which is set at $190 $ meV instead of $180 $ meV. 
 As a result, when $L$ is in the range $[0 , 20[$ nm, the particles propagate through the barrier with the Fermi velocity $v_F$ for armchair directions, but only in the case where $S=0$ for zigzag directions. 
As $L$ is increased, we see that $\tau_{t}/\tau_{0}$ gradually increases to a maximum and then rapidly decays to to a constant value independent of $L$. The Hartman effect and superluminal tunneling can be observed in both strain directions, as seen in \cite{Fattasse}. 
We observe that $S$, $ E $, and $\phi$ have an effect on $\tau_{t}/\tau_{0}$ behavior because it increases as they increase for armchair and zigzag directions. 

\begin{figure}[H]
	\centering
	\subfloat[$S=0$, $S=0.1$, $S=0.15$]{
		\centering
		\includegraphics[width=8cm, height=4.8cm]{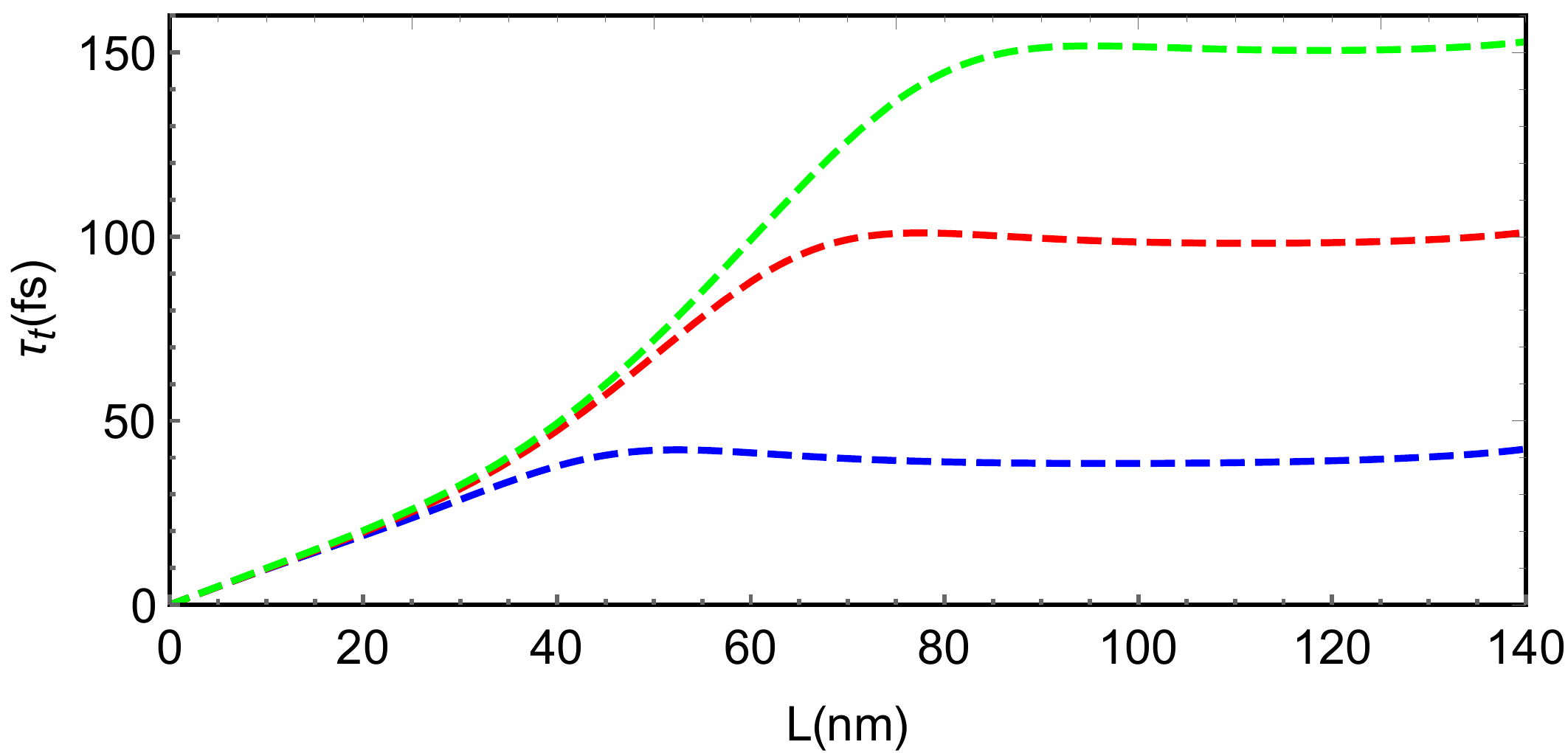}
		\label{fig8a}
	}\hspace{1cm}
\subfloat[$S=0$, $S=0.1$, $S=0.15$]{
		\centering
		\includegraphics[width=8cm, height=4.8cm]{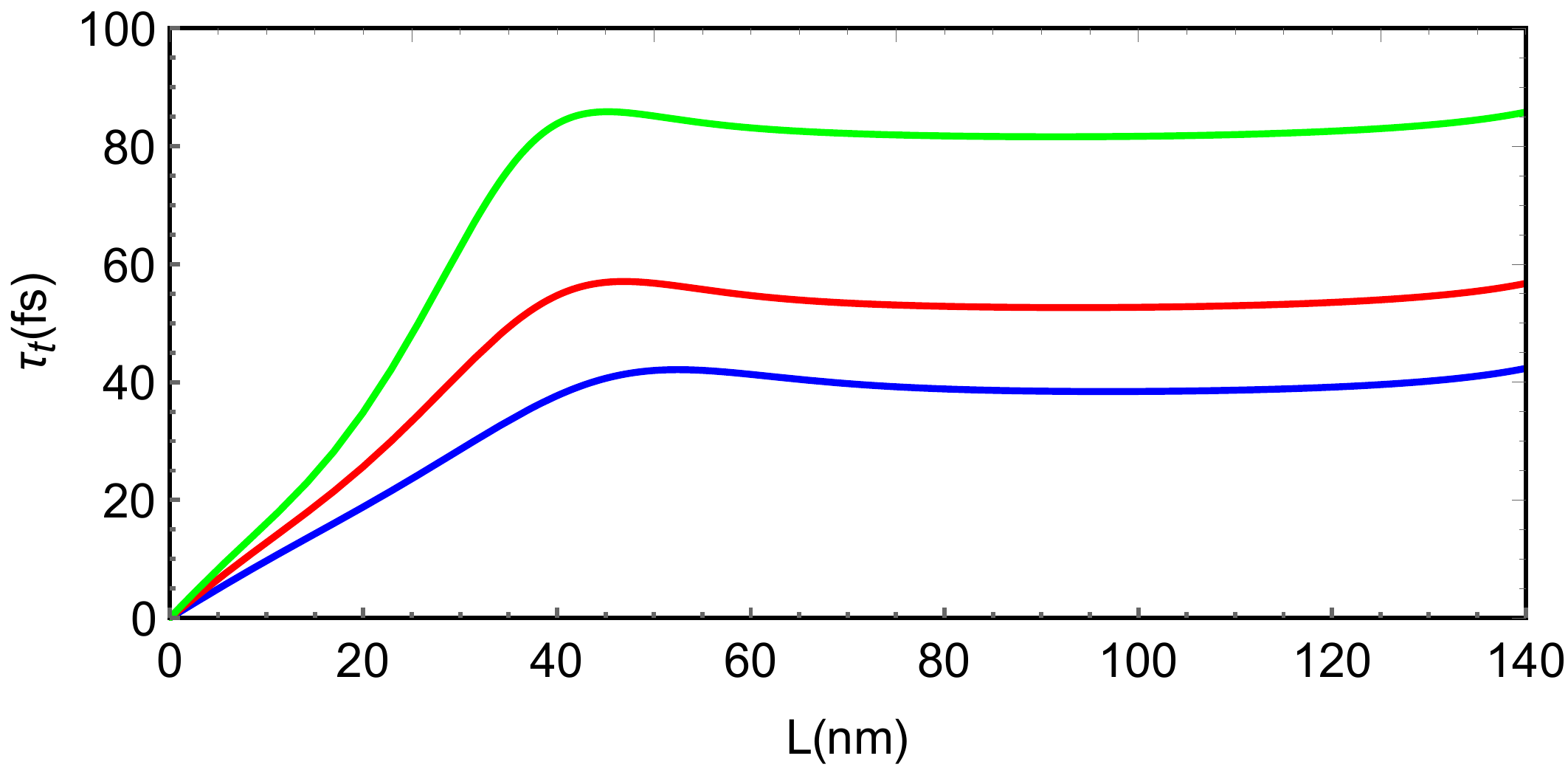}
		\label{fig8b}}\\
	\centering
	\subfloat[$B=0.8$ T, $B=0.9$ T, $B=1$ T]{
		\centering
		\includegraphics[width=8cm, height=4.8cm]{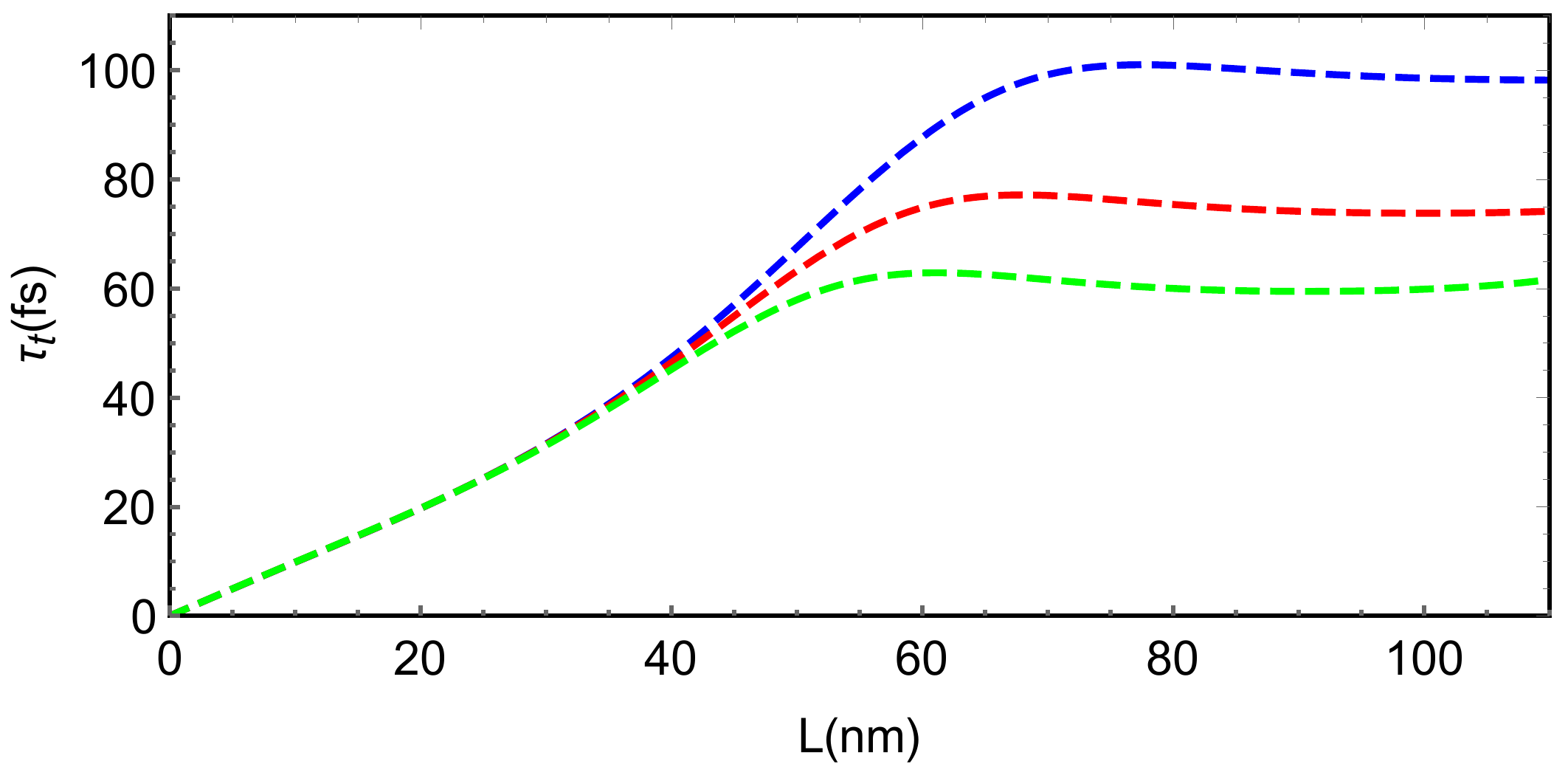}
		\label{fig8c}
	}\hspace{1cm}
\subfloat[$B=0.8$ T, $B=0.9$ T, $B=1$ T]{
		\centering
		\includegraphics[width=8cm, height=4.8cm]{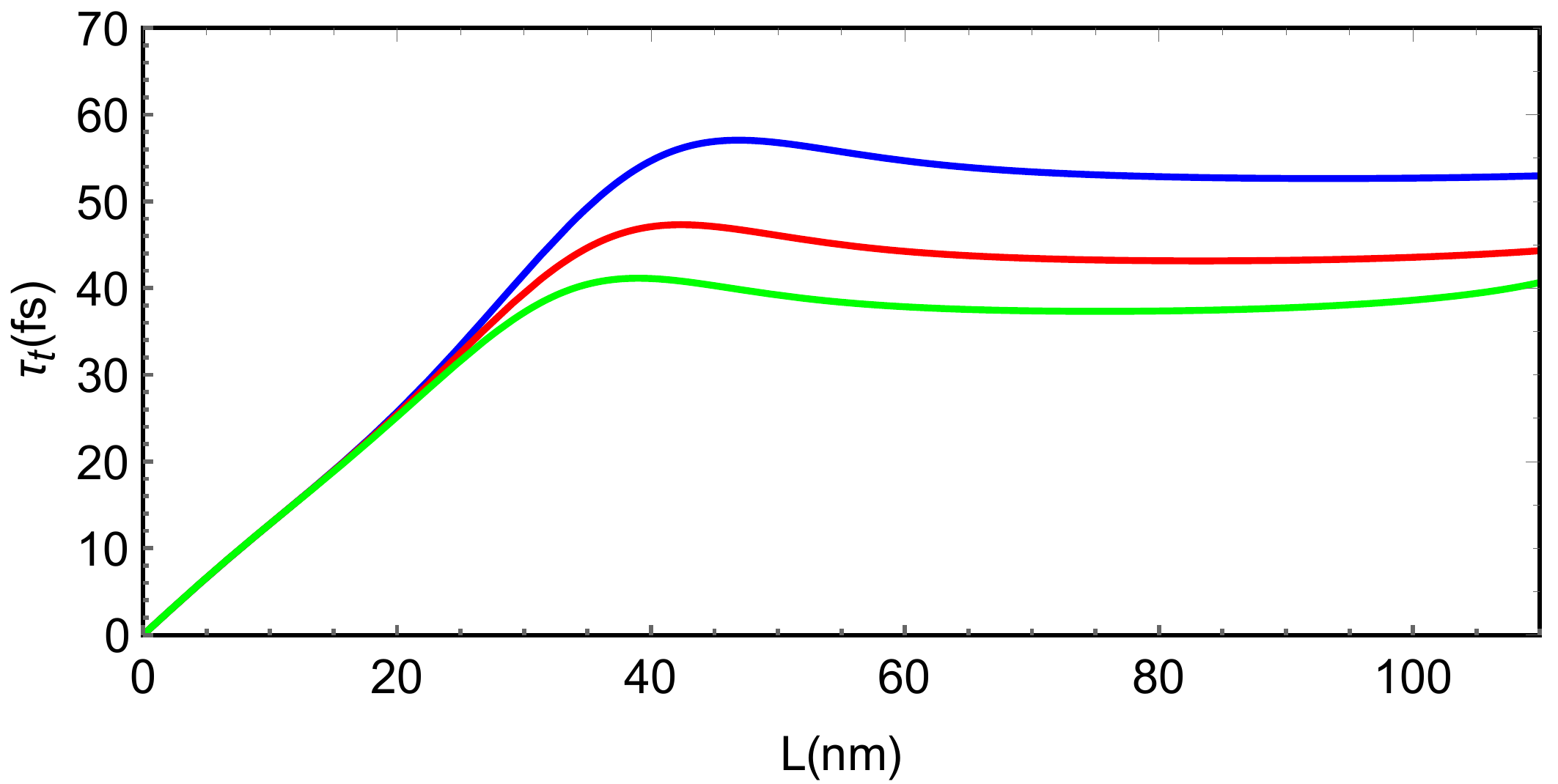}
		\label{fig8d}}\\
	\centering
	\subfloat[$\phi=0^{\circ}$, $\phi=10^{\circ}$, $\phi=15^{\circ}$]{
		\centering
		\includegraphics[width=8cm, height=4.8cm]{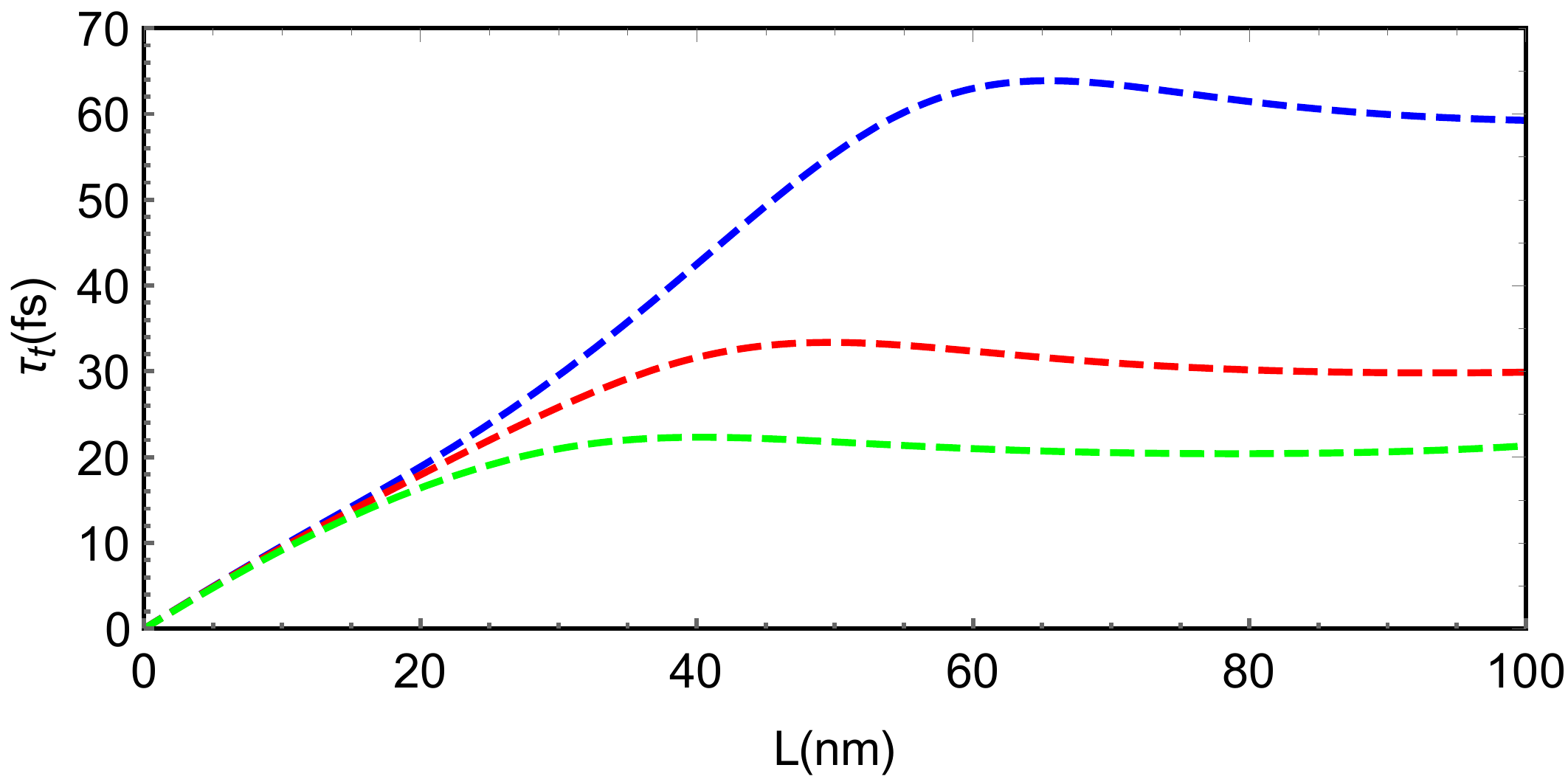}
		\label{fig8e}
	}\hspace{1cm}
\subfloat[$\phi=0^{\circ}$, $\phi=10^{\circ}$, $\phi=15^{\circ}$]{
		\centering
		\includegraphics[width=8cm, height=4.8cm]{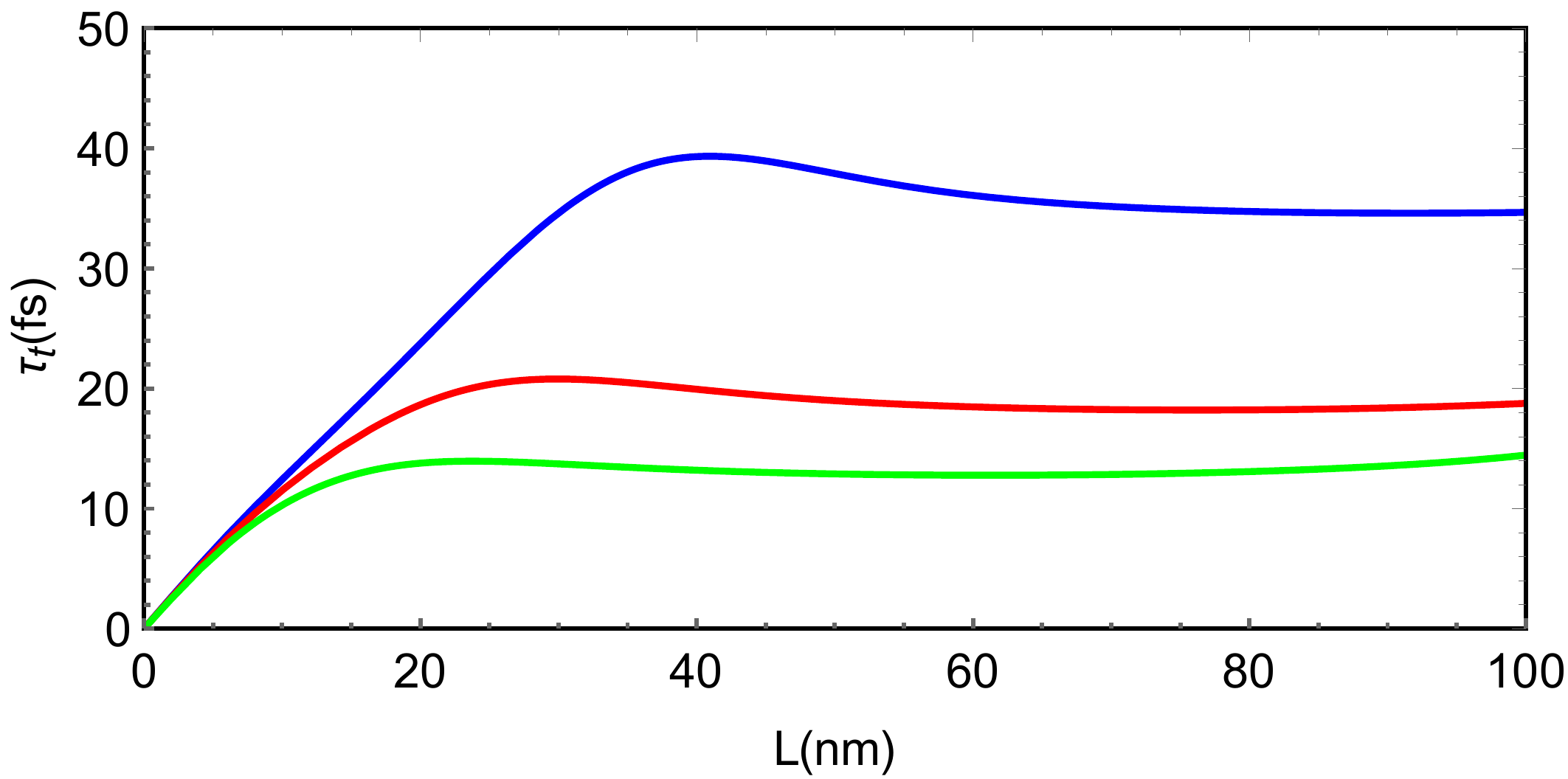}
		\label{fig8f}}
	\caption{{(color online) The group delay time  $\tau_{t}$ versus
			the barrier width $L$ for $V=100$ meV, $E=150$ meV and $\Delta=10$ meV.   \textbf{\color{red}{(a,b)}}:
			$S=0$ (blue), $S=0.1$ (red),  $S=0.15$ (green)
			with $B=0.8 $ T, $\phi=10^{\circ}$.  \textbf{\color{red}{(c,d)}}: 	$B=0.8$ T (blue), $B=0.9$ T (red),  $B=1$ T (green) with $S=0.1 $, $\phi=10$.  \textbf{\color{red}{(e,f)}}: 
			$\phi=0^{\circ}$ (blue), $\phi=10^{\circ}$ (red), $\phi=15^{\circ}$ (green)
			with  $B=0.8$ T, $S=0.1$.  \textbf{\color{red}{(a,c,e)}}: 	Armchair direction
			\textbf{\color{red}{(b,d,f)}}:
			zigzag  direction. }}
	\label{fig8}
\end{figure} 

The influence of the barrier width $L$ on the group delay time  $\tau_ t$ for the armchair (\ref{fig4a},\ref{fig4c},\ref{fig4e}) and zigzag directions  (\ref{fig4b},\ref{fig4d},\ref{fig4f}) is shown in Fig. \ref{fig8}. As the barrier width $L$ is increased, the group delay $\tau_{t}$  saturates at a constant value, demonstrating that the Hartmann effect exists in this tunneling process \cite{YueBan}.
Figs. (\ref{fig8a},\ref{fig8b}) show the effect  of three strain values: $S=0$ (blue), $S=0.1$ (red),  $S=0.15$ (green) with $V=100$ meV, $E=150$~meV, $B=0.8$ T, $\Delta=10$ meV, $\phi=10^{\circ}$. As a result, we see that $\tau_t$ rises when $S$ rises.   
In Figs. (\ref{fig8c},\ref{fig8d})  we consider three values of the magnetic field: $B=0.8$ T (blue), $B=0.9$ T (red),  $B=1$ T (green). Figs. (\ref{fig8e},\ref{fig8f}) present three values of the incident angle: $\phi=0^{\circ}$ (blue), $\phi=10^{\circ}$ (red), $\phi=15^{\circ}$ (green). We see that as $ B $ and $\phi$ increase the group delay time, $\tau_t$ decreases, which is consistent with the result obtained in \cite{YueBan}.

    \section{Conclusion}
    
    For transmitted Dirac fermions in gapped graphene through a magnetic barrier, we have theoretically examined the effects of strain applied along armchair and zigzag directions on the GH shifts, the group delay time, and the Hartman effect. This work is based on a two-dimensional stationary phase approach.   The second of the three zones that make up the current system is the one that is subject to the strain effect and the external magnetic field. By resolving the Dirac equation, we were able to derive the energy spectrum and related eigenspinors. We were able to create a transfer matrix linking the propagation amplitudes in the input and output regions thanks to the continuity of the eigenspinors at each interface separating two adjacent regions. For zigzag strain and armchair strain orientations, we estimated the transmission probabilities along with the GH shifts and group delay time.

The group delay time $\tau_{t}/\tau_{0}$ was then quantitatively shown in transmission plots versus potential $V$, magnetic field $B$, energy gap $\Delta$, and barrier width $L$ under suitable conditions. The armchair and zigzag direction strain charts for the group delay time revealed that transmission resonances can increase it in the propagating mode for small values of $V$, $B$, $\Delta$, and $L$. 
When these values are increased, the group delay decreases and becomes less than $L/v_F$, and the group delay can be changed from subluminal to superluminal by adjusting various physical parameters. Furthermore, when these parameters are null, particles pass through the barrier at the Fermi velocity of $v_F$ regardless of the value of $S$ for armchair directions and $S=0$ for zigzag directions. When $S\ne0$, however, the particles move through the barrier at a velocity less than Fermi's $v_F$. Moreover, for the zigzag direction, the oscillation of group delay time increases compared to strainless graphene and armchair directions. Besides, the Hartman effect can appear for all incident angle and for both strain directions, in contrast to results presented in \cite{YueBan}, where the Hartman effect can be observed only for the normal incident angle $\phi=0$.


\end{document}